\documentclass[12pt,preprint]{emulateapj}
\usepackage{natbib}
\usepackage{color}

\newcommand{\be}{\begin{equation}}
\newcommand{\ee}{\end{equation}}
\newcommand{\ba}{\begin{eqnarray}}
\newcommand{\ea}{\end{eqnarray}}

\begin{document}

\title{HerMES: A Statistical Measurement of the Redshift Distribution of Herschel-SPIRE Sources Using the Cross-correlation Technique}



\author{K.~Mitchell-Wynne\altaffilmark{1},
A.~Cooray\altaffilmark{1,2},
Y.~Gong\altaffilmark{1},
M. B\'{e}thermin\altaffilmark{3,4},
J.~Bock\altaffilmark{2,5},
A.~Franceschini\altaffilmark{6},  
J.~Glenn\altaffilmark{7,8},
M.~Griffin\altaffilmark{9},
M.~Halpern\altaffilmark{10},
L.~Marchetti \altaffilmark{6},
S.~J.~Oliver\altaffilmark{11},
M.J.~Page\altaffilmark{12},
I.~P{'e}rez-Fournon\altaffilmark{13,14},
B.~Schulz\altaffilmark{2},
D.~Scott\altaffilmark{10},
J.~Smidt\altaffilmark{1},
A.~Smith\altaffilmark{11},
M.~Vaccari\altaffilmark{6,15}, 
L.~Vigroux\altaffilmark{16}, 
L.~Wang\altaffilmark{11},
J.~L.~Wardlow\altaffilmark{1},
M.~Zemcov\altaffilmark{2,5}}

\altaffiltext{1}{Dept. of Physics \& Astronomy, University of
California, Irvine, CA 92697}
\altaffiltext{2}{California Institute of Technology, 1200 E.
California Blvd., Pasadena, CA 91125}
\altaffiltext{3}{Institut d'Astrophysique Spatiale (IAS), bat121, F-91405 Orsay,
France; Universit«e Paris-Sud 11 and CNRS (UMR8617)}
\altaffiltext{4}{Laboratoire AIM-Paris-Saclay, CEA/DSM/Irfu - CNRS -
Universit´e Paris Diderot, CE-Saclay, pt courrier 131, F-91191
Gif-sur-Yvette, France}
\altaffiltext{5}{Jet Propulsion Laboratory, 4800 Oak Grove Drive,
Pasadena, CA 91109}
\altaffiltext{6}{Dipartimento di Astronomia, Universit\`{a} di Padova,
vicolo Osservatorio, 3, 35122 Padova, Italy}
\altaffiltext{7}{Center for Astrophysics and Space Astronomy 389-UCB, University of Colorado, Boulder, CO 80309}
\altaffiltext{8}{Dept. of Astrophysical and Planetary Sciences, CASA
389-UCB, University of Colorado, Boulder, CO 80309}
\altaffiltext{9}{School of Physics and Astronomy, Cardiff University,
Queens Buildings, The Parade, Cardiff CF24 3AA, UK}
\altaffiltext{10}{Department of Physics \& Astronomy, University of
British Columbia, 6224 Agricultural Road, Vancouver, BC V6T~1Z1,
Canada}
\altaffiltext{11}{Astronomy Centre, Dept. of Physics \& Astronomy,
University of Sussex, Brighton BN1 9QH, UK}
\altaffiltext{12}{Mullard Space Science Laboratory, University College
London, Holmbury St. Mary, Dorking, Surrey RH5 6NT, UK}
\altaffiltext{13}{Instituto de Astrof\'{i}sica de Canarias (IAC),
E-38200 La Laguna, Tenerife, Spain}
\altaffiltext{14}{Departamento de Astrof\'{i}sica, Universidad de La
Laguna (ULL), E-38205 La Laguna, Tenerife, Spain}
\altaffiltext{15}{Astrophysics Group, Physics Department, 
University of the Western Cape, Private Bag X17, 7535, Bellville, Cape Town, South Africa}
\altaffiltext{16}{Institut d'Astrophysique de Paris, UMR 7095, CNRS,
UPMC Univ. Paris 06, 98bis boulevard Arago, F-75014 Paris, France}

\begin{abstract}
The wide-area imaging surveys with the {\it Herschel} Space Observatory at sub-mm wavelengths
have now resulted in catalogs of order one hundred thousand dusty, star-burst galaxies.
These galaxies capture an important phase of galaxy formation and evolution, 
but unfortunately the redshift distribution of these galaxies, $N(z)$, 
is still mostly uncertain due to limitations associated with counterpart identification at optical 
wavelengths and spectroscopic follow-up.
We make a statistical estimate of $N(z)$ using a clustering analysis 
of sub-mm galaxies detected at each of 250, 350 and 500$\,\mu$m  from the {\it Herschel} 
Multi-tiered Extragalactic Survey (HerMES)
centered on the Bo\"{o}tes field. We cross-correlate {\it Herschel} galaxies against galaxy 
samples at optical and near-IR wavelengths from the Sloan Digital Sky Survey (SDSS),
the NOAO Deep Wide Field Survey (NDWFS) and the {\it Spitzer} Deep Wide Field Survey (SDWFS).  
We create optical and near-IR galaxy samples
based on their photometric or spectroscopic redshift distributions and test 
the accuracy of those redshift distributions with similar galaxy
samples defined with catalogs from the Cosmological Evolution 
Survey (COSMOS), which has superior spectroscopic coverage.
We model the clustering auto- and cross-correlations of {\it Herschel} 
and optical/IR galaxy samples to estimate $N(z)$ and clustering bias
factors. The $S_{350} > 20\,$mJy galaxies have a bias factor varying 
with redshift as $b(z)=1.0^{+1.0}_{-0.5}(1+z)^{1.2^{+0.3}_{-0.7}}$.
This bias and the redshift dependence is broadly in agreement with 
galaxies that occupy dark matter halos of mass
in the range of 10$^{12}$ to 10$^{13}$ M$_{\sun}$. We find that galaxy selections in
all three SPIRE bands share a similar average redshift, with $\langle z \rangle = 1.8 \pm 0.2$ 
for 250$\,\mu$m selected samples, and $\langle z \rangle = 1.9 \pm 0.2$ for both 350 and 
500$\,\mu$m samples, while their
distributions behave differently. For 250$\,\mu$m selected
galaxies we find the a larger number of sources with $z \le 1$ when compared with the
subsequent two SPIRE bands, with 350 and 500$\,\mu$m selected SPIRE samples having
peaks in $N(z)$ at progressively higher redshifts. We compare our clustering-based $N(z)$ results
to sub-mm galaxy model predictions in the literature,  and with an estimate 
of $N(z)$ using a stacking analysis of COSMOS 24$\,\mu$m detections. 
\end{abstract}


\maketitle

\section{Introduction}

The properties of the dusty, star-forming galaxies detected by the Spectral and 
Photometric Imaging Receiver (SPIRE; Griffin et al. 2010) 
aboard the {\it Herschel Space Observatory}\footnote{{\it Herschel} 
is an ESA space observatory with science instruments provided by 
European-led Principal Investigator consortia and with important participation from NASA.} 
(Pilbratt et al. 2010) at sub-mm wavelengths provide important clues to the 
nature of dusty star-formation and the role of galaxy mergers in triggering such 
star-formation in distant galaxies.  However, the redshift distribution of these 
galaxies has yet to be determined observationally.
The low spatial resolution of {\it Herschel}-SPIRE observations 
complicate the identification of counterparts at optical and near-IR
wavelengths.  Moreover, the optical emission from these 
star-bursting galaxies is highly extinct and could potentially bias
optical spectroscopy observations to low-redshift bright 
galaxies. Instead of optical or IR spectroscopy, mm and sub-mm wave
spectroscopy can be pursued targeting fine-structure 
and molecular lines such as CO and [CII]. Such measurements, unfortunately, 
are currently limited to a handful of the brightest sources -- 
mostly the rarely lensed sub-mm galaxies (e.g., Lupu et al. 2010; Scott et al. 2011;
Riechers et al. 2011; Harris et al. 2012), as existing instrumental capabilities do not 
allow large CO or [CII] surveys of typical sub-mm galaxies. 

There have been a few other approaches to obtain the $N(z)$ of sources at these wavelengths.
A statistical approach based on photometry alone, using SPIRE colors, was 
considered in Amblard et al. (2010; see also Lapi et al. 2011), 
but such techniques are subject to uncertainties on the assumed 
spectral energy distribution (SED) of the  galaxies at sub-mm 
wavelengths. These generally involve isothermal SED models, where the redshift
distribution is degenerate with the assumed dust temperature distribution.
Marsden et al. \citep{Mars09} employed stacking methods to effectively measure the CIB as a
function of redshift, and B\'{e}thermin et al. \citep{Beth12} have recently measured
deep source counts as a function of redshift, also via stacking, which is
compared to with results of this paper.

Here we pursue a second statistical approach to measure the SPIRE 
galaxy redshift distribution using the spatial clustering of the sub-mm 
population relative to clustering of galaxies with an a priori known 
redshift distribution (Schneider et al. 2006; Newman 2008; Zhang et al. 2010). 
The unknown sub-mm redshift distribution can be estimated 
via the strength of its cross-correlation relative to galaxy samples of known redshifts.
Modeling also requires that the clustering bias factors of all 
galaxies be determined jointly through a combination of auto and
cross-correlation functions. The key advantage of this technique is that 
it does not require cross-identification of SPIRE sources in optical and IR catalogs.

For this study we make use of data from the \textit{Herschel} Multi-tiered 
Extragalactic Survey (HerMES; Oliver et al. 2012), which 
mapped a large number of well-known fields with existing multi-wavelength 
ancillary data using SPIRE.  To cross-correlate against 
SPIRE-selected galaxies, we make use of near-IR selected galaxy samples
from {\it Spitzer} observations based on the 1.6$\,\mu$m ``bump'', 
which has long been established as a redshift indicator 
(Sawicki 2002; Simpson \& Eisenhardt 1999; Wright \& Fazio 1994). 
The bump results from the fact that the $\text{H}^{-}$ absorption in stellar 
atmospheres is minimally opaque at 1.6$\,\mu$m. This leads 
to a bump in the spectral energy distributions of cool stars at 1.6$\,\mu$m \citep{Joh88} that
is nearly ubiquitous in galaxy spectra. 
For $z > 0$, the wavelength at which the bump in the SED peaks allows for a redshift determination
based on the colors in IRAC channels between 3.6 and 8$\,\mu$m, and covering the redshift range of 1 to 2.5. 
We complement these ``bump'' galaxy samples with a 24$\,\mu$m 
and an $R$-band based sample of dust obscured galaxies, which has a redshift distribution that peaks
around $z \sim 2.3$ (Dey et al. 2008). We also make use of 
optical-selected galaxy samples with SDSS spectroscopic and photometric redshifts out to about 0.7.

The paper is organized as follows. In Section~2 we describe source selection in 
SPIRE and galaxy sample selection with IRAC and MIPS bands, 
complemented with optical data to remove outliers. In Section~3 
we describe the redshift distribution of the galaxy samples used for the 
cross-correlation analysis, and in Section~4 we describe the cross-clustering measurements. Fitting results
are presented in Section~5.  In Section~6 we present $N(z)$ and 
$b(z)$, and discuss our results. We assume a flat-$\Lambda$CDM 
cosmological model and fix the cosmological parameters to the 
best-fit values of $\Omega_{\text{m}} = 0.27$, $\Omega_b=0.046$, $\sigma_8=0.81$, $n_s=0.96$
and $h=0.71$  \citep{WMAP7} when performing MCMC model fits. 

\section{SPIRE Source and Galaxy Sample Selection}

\subsection{{\it Herschel}-SPIRE sample}

The HerMES SPIRE source catalogs used for this paper 
come from a combined analysis involving both a
direct source extraction and an attempt to account for 
blending at 350 and 500 $\mu$m 
wavelengths given the positions of 250 $\mu$m 
detections (Wang et al.~in prep). The method updates the source extraction
pursued by HerMES at each of the three SPIRE bands 
independently that ignored issues associated 
with blending at longer wavelengths (Smith et al. 2011). 

In order to maximize the overlap with multi-wavelength data, we concentrate our study 
on the  Bo\"{o}tes field with HerMES SPIRE data covering
12.5 deg$^{2}$. The field has been imaged with {\it Spitzer} IRAC
as part of the  {\it Spitzer} Deep Wide Field Survey (SDWFS; Ashby et al.~2009) 
and from the ground with optical to near-IR observations as part of the 
NOAO Deep Wide Field Survey (NDWFS; Jannuzi and Dey 1999), with coverage also 
provided by the Sloan Digital Sky Survey (SDSS; Abazajian et al. 2009).  

For this study we selected SPIRE sources with a flux density
greater than 20 mJy in the B\"{o}otes field. We find in 
excess of 22\,000 sources in each of the SPIRE 
bands covering the entire 12.5 deg$^{2}$ of
SPIRE observations,  while 3775, 3243 and 958 galaxies 
at 250 $\mu$m, 350 $\mu$m and 500 $\mu$m,  respectively, 
were used in the cross-correlation study -- an area covering 
6.7 deg$^{2}$, where various ancillary data best overlap.

At 20 mJy, the SPIRE catalogs are $\sim$ 30\% complete at each of the three wavelengths. The 90\% completeness
of the catalogs is at  a flux density of about 55 mJy (Wang et al. in prep). 
At such a high flux density level, the number of  SPIRE sources in the area overlapping with ancillary data is down by at least factor of 8 and the resulting low 
surface density does not allow useful constraints on the redshift distribution. We note some caution in interpreting
our results with models due to the incompleteness. We are unable to correct for it through simulations 
due to the lack of a priori information on the redshift distribution of missing sources. It is unlikely, however,
that the redshift distributions presented here are biased due to catalog incompleteness since the source detection algorithm is
primarily based on the flux density and not the individual redshifts of SPIRE sources.

\subsection{IRAC Sample Selection and Star-Galaxy Separation}

Using the SDWFS data combined with ground-based {\it K}-band data from NDWFS, 
we generated three different catalogs of 1.6$\,\mu$m-bump sources based on the IRAC 
channel where the SED peaks.
These three samples are as follows: bump-1 with
a peak in the 3.6$\,\mu$m channel ($0.5 \lesssim z \lesssim  1.5$); bump-2 peaking in the 
4.5$\,\mu$m channel ($0.8 \lesssim z \lesssim 2.2$); and bump-3 with a peak at 5.8$\,\mu$m 
($1.5 \lesssim z \lesssim 3.0$).  Using the photometric redshifts 
computed via a template fitting method (Csabai et al. 2003) in 
the SDSS DR7 catalog, we also constructed two separate  redshift distributions 
with peaks at $z \sim 0.3$ and 0.7.

In order to establish catalogs of bump-1 to bump-3 galaxy populations we first had to 
remove stars and other contaminants from our optical and IR catalogs. 
This was done using a combination of infrared and optical data. We used the SDWFS 
four-epoch stacked catalog (Ashby et al.~2009) which contains all 
sources detected in the first channel of IRAC at or above 5$\sigma$. 
This catalog was matched with the NDWFS third data release catalog and the SDSS 
catalog, using a 2.5$''$ matching radius. For sources with multiple matches ($<$3\%), 
magnitudes from the NDWFS catalog were then 
compared with the 3.6$\,\mu$m magnitude and entries with the most similar values were kept.

\begin{figure}
  {\includegraphics[scale=.5]{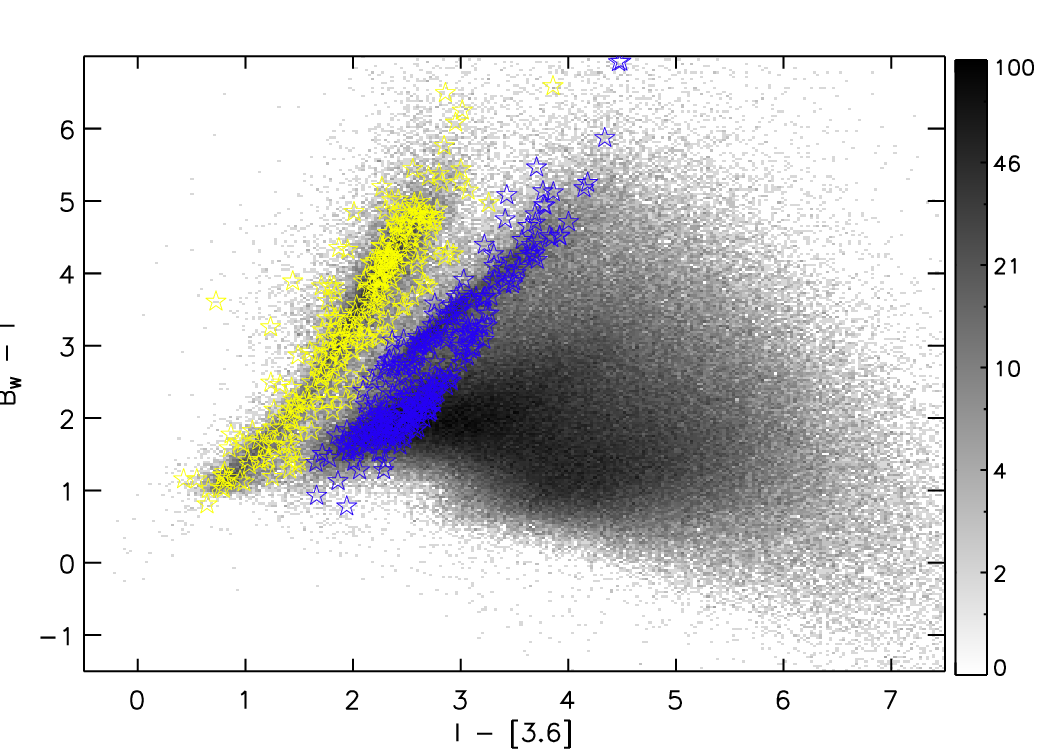}
  \includegraphics[scale=.5]{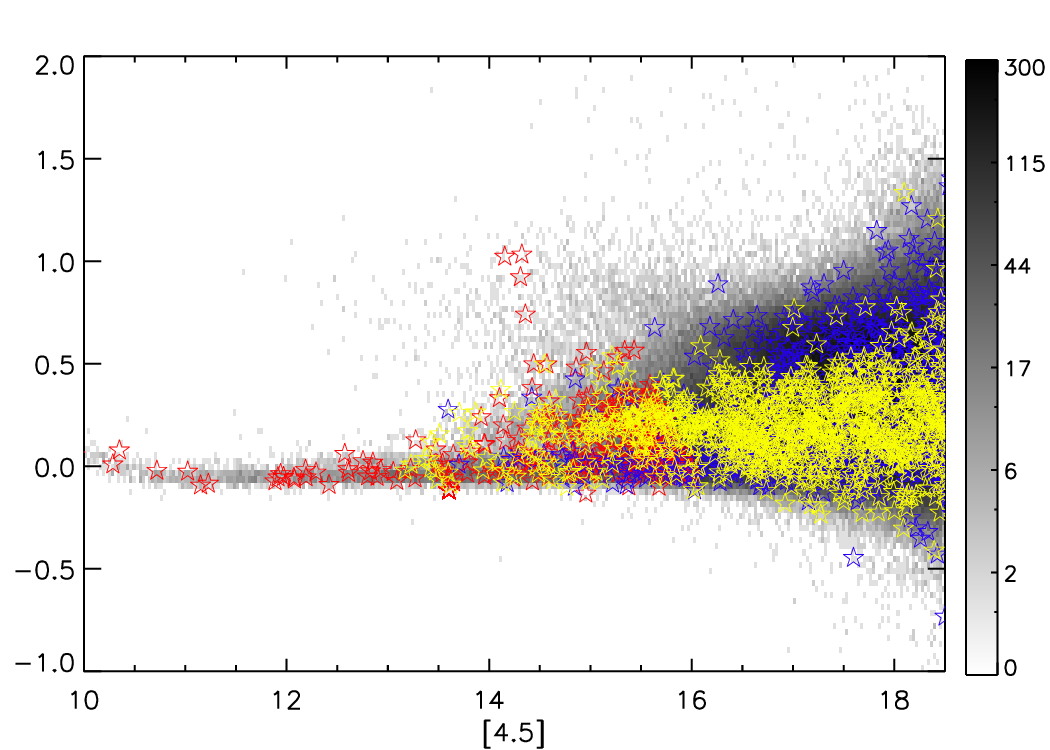}}
  \caption{Color-color (top; $B_{\text{w}}-I$ vs. $I-[3.6]$) and color-magnitude (bottom; $[3.6]-[4.5]$ vs. [4.5]) 
   density plots of sources (both stars and galaxies) in our Bo\"otes field 
   catalogs. Red stars indicate the 3.6 $\mu$m magnitude selection for stars, 
  while giants are depicted as yellow stars, and the \textit{BIK} 
  sequence as blue stars (see Section~2.2 for details). 
 Only a fraction of the sources identified as stars 
  are plotted to avoid over-crowding in the plots. Black points are the galaxies.}
  \label{fig:star_reduc}
\end{figure}

Stars and spurious sources were removed from the resulting 
merged catalog using various techniques. Vega magnitudes and 6$''$ 
diameter aperture photometry are used throughout the 
star-galaxy separation unless otherwise noted. We employed a three 
stage process to remove stars from our catalog. An initial selection of 
sources with [3.6] $<$ 16 were identified as stars, where [3.6] represents the
vega magnitude at 3.6 $\,\mu$m. Using the combination of 
optical and IRAC photometry we further classify sources as stars that either 
satisfy ($B_{\text{w}} - I) > 2(I - [3.6]) - 1.65$, or $-1.65 > 
(B_{\text{w}} - I) - 2(I - [3.6]) > -3.35$ \citep{Eis04}. The former criterion 
defines a sequence of $BIK$ stars \citep{Hua97}, and the 
latter a sequence of giant stars (\cite{Joh66}; \cite{Bes88}). Lastly, for 
IRAC sources without optical counterparts, we used a binning 
method that involved only the IRAC bands \citep{Wad07}. Three flux density bins were defined 
with the criteria [3.6] $\le$ 19.5, 19.5 $<$ [4.5] $\le$ 20.0 
and [4.5] $\le$ 23, with color cuts [3.6] - [4.5] $<$ $-0.35$, $-0.30$ 
and $-0.25$, respectively; all sources satisfying these criteria were assumed 
to be stars. The results of these extractions are shown in Fig. \ref{fig:star_reduc}.

\subsection{Bump, DOG and SDSS Selections}

From our resulting merged and star-subtracted catalog, we invoked 
simple color constraints to classify three different types of
bump sources and dust-obscured galaxies (DOGs; Dey et al. 2008), using 4"
aperture diameter photometry. Bump-1, bump-2 and bump-3 sources each display excess
emission in IRAC channels 1, 2 and 3, respectively. Bump-1 sources 
were selected using the criteria $K - [3.6] > 0.1$ and $[3.6] - [4.5] < 0$; 
bump-2 with $K - [3.6] > 0$, $[3.6] - [4.5] > 0$ and $[4.5] - [5.8] < 0$; 
bump-3 with $[3.6] - [4.5] > 0$, $[4.5] - [5.8] > 0$ and $[5.8] - [8] < 0$. 
The number of bump-1 sources in the SDWFS catalogs were found to be $\sim$ 
1.3 $\times 10^{4}$ at the 5$\sigma$ detection limit of the 3.6$\,\mu$m channel of the IRAC instrument. 
Bump-2 source identification yielded 6.5 $\times 10^{3}$ galaxies, while the bump-3 catalog contains 4 $\times 10^{3}$ galaxies.

We also make use of a sample of dust obscured galaxies, 
selected with 24 $\mu$m {\it Spitzer}-MIPS and optical 
{\it R}-band data to have extreme red colors from dust obstruction, with 
S$_{24}$/S$_{R} >$ 1000 (where S$_{24}$ is the 24$\,\mu$m flux density), or equivalently $R - [24] \ge 14$
and S$_{24} \ge$ 0.3 mJy ($\approx 6 \sigma$; Dey 2009). We found that a total of
2838 galaxies satisfied the selection criteria. Based on spectroscopic follow-up,
they are now known to have a  mean redshift around $z \sim 2.3$ \citep{Dey09}.
We make use of the full, broad redshift distribution for this sample, 
spanning the range of $0.5 < z < 3.5$, with a peak
around $z \sim 2$, found from a similar identification 
of DOGs in the COSMOS field (see Section~3 and Fig.~2) for the present analysis.
 These dust-obscured galaxies have  been suggested to be 
 an intermediary phase of the evolution of quasi-stellar objects from gas-rich mergers \citep{Dey09}. 
They have also been shown to be strongly clustered and are believed to be progenitors 
of massive ($3-6L_{*}$) galaxies at low redshift \citep{Bro08}. 

Finally, to cover the redshift range of $ 0 < z < 0.7$ 
efficiently we also selected optical galaxies from SDSS. These sources have photometric 
redshifts, individually determined with SED fits to SDSS photometry, in the above range.
We make use of 8\,000 SDSS galaxies  and we 
consider two sub-samples peaking at $z \sim 0.2$  and 0.5 with roughly equal
numbers. The first of these two sub-samples is obtained by selecting sources which
obey $2.6 < B - I < 3$ and $-0.8 < I - R < 0.1$, while the second
selection obeys $B - I >$ 4 and $-0.9 < I - R < 0$. These six galaxy samples (3 bump catalogs, DOGs, 
and two SDSS samples) provide adequate redshift coverage over the range of $0 < z < 3$.

\begin{figure}[h!]
  {\includegraphics[scale=.5]{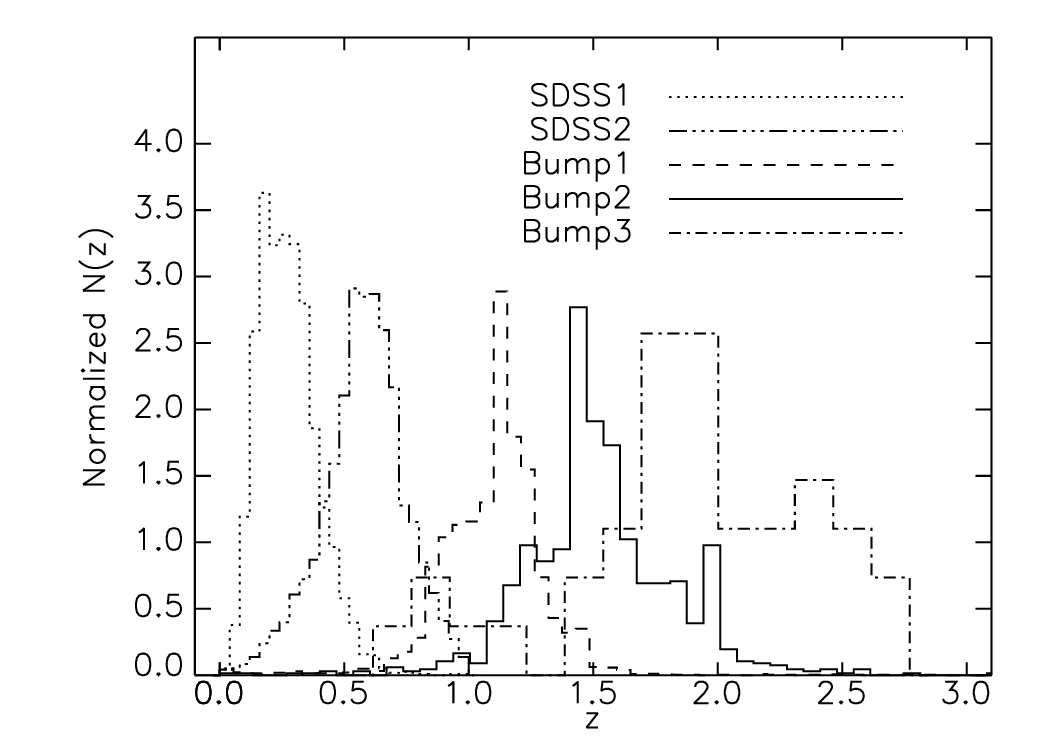}}
  {\includegraphics[scale=.5]{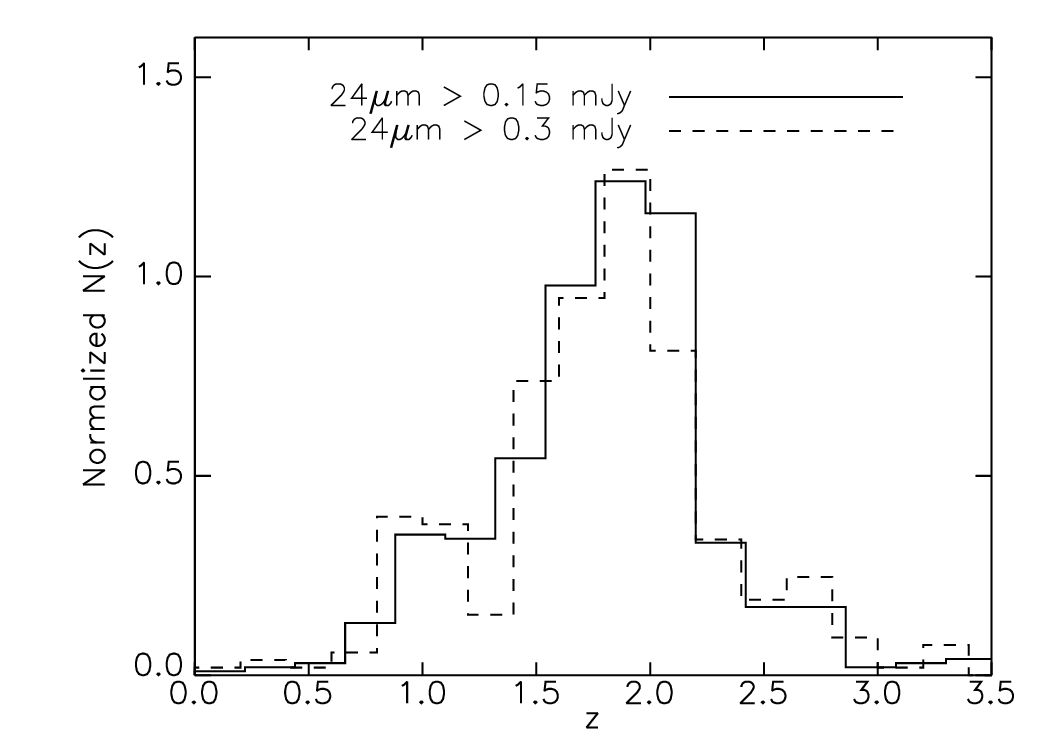}}
  \caption{$N(z)$ distribution obtained from SDSS photometric redshifts in the Bo\"{o}tes field 
  as well as bump-1 to -3 (top plot) and DOGs  (bottom plot)
  redshift distributions from the COSMOS field. We assume the latter four redshift distributions 
  measured directly in COSMOS are also applicable for galaxy samples
  under the same color selection criteria in the Bo\"{o}tes field.}
  \label{fig:hists}
\end{figure}

\section{COSMOS photo-z and spec-z}

While we are able to generate large samples of galaxies to cross-correlate against the 
SPIRE catalogs of the Bo\"{o}tes field, the existing spectroscopic
and photometric redshift information in the  Bo\"{o}tes field is not 
adequate to establish the redshift distributions of the {\it Spitzer} galaxy samples.
For that we turn to data in the Cosmological Evolution Survey (COSMOS; Scoville et al. 2006; Capak et al. 2007) 
where we can select similar galaxy samples as in the Bo\"{o}tes field, using the same depth and color
criteria. For those galaxies we are able to use either the existing spectroscopic 
or the photometric redshifts from the public COSMOS catalog
\citep{Ilb09}. We assume that the redshift distributions for galaxy samples in 
COSMOS is the same as for Bo\"{o}tes when interpreting 
clustering measurements from the wider Bo\"{o}tes area that overlaps with SPIRE.

The COSMOS field has extensive photometric redshift measurements 
 over 2 deg$^2$ using 30 broad, intermediate and narrow band 
filters from space-based telescopes ({\it Hubble, Spitzer, GALEX, XMM, Chandra}) 
and ground-based telescopes (Subaru, VLA, ESO-VLT, UKIRT, NOAO, 
CFHT, and others). We used a public COSMOS source catalog containing 
$\sim 10^{4}$ spectroscopic redshifts, and $\sim 3 \times 10^{5}$ photometric 
redshifts that were computed using a $\chi^{2}$ galaxy template-fitting technique \citep{Ilb09}.
In both the B\"{o}otes and COSMOS fields, we imposed the same 
brightness thresholds in the selection bands: for IRAC channel 1 and {\it K}-band
we imposed a 5$\sigma$ detection limit. We require that each of the galaxies detected at above 5$\sigma$ is
also detected in IRAC Channcel 2 to 4. In those cases, however, we considered a source that has a flux density above 50\%
completeness level to be considered as detected, while sources with flux densities below 50\% completeness level of
each of the three channels were dropped from the final catalog. This selection process was chosen to ensure
that we are probing equal depths between the two fields, COSMOS and  B\"{o}otes.

Using the same selection methods described above, we were able to obtain 
$N(z)$ measurements for each of our different source classifications 
from the COSMOS source catalog (see Fig.~\ref{fig:hists}). Note that the galaxy
type selections are mutually exclusive, so there are no overlapping sources between
different samples. All of the bump sources have well 
defined redshift distributions, and the DOG distribution agrees 
well with those in the literature (see figure~1 in Dey 2009). We identified 384 sources 
as DOGs in COSMOS, with $S_{24} > 0.3\,\mu$Jy, and 683 with $S_{24} > 0.15\,\mu$Jy, 
a number density consistent with statistics of the DOG population 
in other fields \citep{Bro08}.

\section{Angular cross-correlation and covariance matrix}

To obtain the redshift distributions of SPIRE sources, 
we first cross-correlate the SDSS-selected sample and bumps and DOGs 
from the Bo\"{o}tes  field, against sub-mm 
sources in each SPIRE band, from arcminute to degree angular scales. 
We also measure the auto-correlation functions of the galaxy and SPIRE 
samples, as these are needed to model the clustering strength and to 
extract the unknown redshift distribution. 

We use a bootstrap method to establish the covariance matrix for  each of the cross- or auto-correlation 
functions, as an accounting of the covariance is needed to properly model the
clustering measurements. We do this by selecting
 200 separate catalogs from the original SPIRE data by removing about 5\% of the sources randomly.
We measure the auto and cross-correlations with each of those catalogs and build the mean auto and cross-correlation functions,
the variance from the scatter, and the covariance from the correlations between the measured auto and cross-correlation functions.

The angular cross-correlation function is modeled  analytically using the COSMOS redshift distribution of the bump-1, bump-2, 
bump-3 and DOGs, while for the SDSS galaxy samples we make use of the public 
photometric redshifts from SDSS DR-7. For simplicity we bin the unknown SPIRE redshift distribution from $z = 0$ to 4
in 5 bins in redshift. To extract the best-fit values and uncertainties in the redshift 
distribution bins, and  the other parameters in the analytical model, we make use 
of a likelihood fitting technique based on the Markov Chain Monte Carlo 
(MCMC) method. 

In this section, we first discuss the method of modeling the angular 
cross-correlation $w_{\rm cross}$ using the redshift distribution of 
galaxies and the linear matter power spectrum. Then we describe 
the measurement of the $w_{\rm cross}$ functions as well as the 
covariance matrix from the galaxy samples. 

\subsection{Modeling the angular cross-correlation}

The angular cross-correlation function $w_{\rm cross}$ for  two galaxy samples
is defined by
\be
w_{\rm cross}(\theta) = \langle \delta n_1(\hat{\phi})\delta n_2(\hat{\phi'}) \rangle,
\ee
where $\delta n_i(\hat{\phi})=(n_i(\hat{\phi})-\bar{n}_i)/ \bar{n}_i$,
and $n_i(\hat{\phi})$ is the number density of galaxies observed 
in direction $\hat{\phi}$ in the sky ($\theta=\phi-\phi'$), and $\bar{n}_i$ is the mean
number density of the galaxy sample $i$. $\delta n_i$ can be
decomposed into two terms--one term from the real clustering of galaxies and a 
second term caused by lensing magnification. Here we ignore the few percent
contribution from lensing \citep{Wang11} and only consider the clustering term, which is 
\ba
w_{\text{gg}}(\theta) = &b_1&b_2\int_0^{\chi_{\text{H}}} d\chi N_1(\chi)N_2(\chi) \\ \nonumber
         &\times& \int_0^{\infty}\frac{k}{2\pi}P(\chi,k)J_0[kr(\chi)\theta],
\ea
where $b_1$ and $b_2$, $N_1(\chi)$ and $N_2(\chi)$ are the galaxy bias and
the normalized radial distribution for the two galaxy samples, respectively.
$P(\chi,k)$ is the power spectrum of the dark matter, $J_0(x)=\rm sin({\it x)/x}$ 
is the zero-order Bessel function, and $\chi$ and $r(\chi)$
are the radial comoving distance and the comoving angular diameter
distance respectively ($r(\chi)=\chi$ in flat space).
$\chi_{\text{H}}$ denotes the radial distance to the horizon, or Hubble length.
Note that $w_{\text{gg}}(\theta)$ will be zero if the positions of the two galaxy samples do not overlap
with eachother.

When modeling the measurements, we make use of the linear theory power 
spectrum to describe $P(\chi,k)$ and focus only on
modeling the measurements over the angular scales of 6$'$, and 
above where the clustering is in the linear regime (Cooray et al. 2000).
At these large angular scales, the 1-halo term makes less than 
a 1\% correction to the correlation function and can be safely ignored.

\subsection{The measurement and covariance matrix of the angular cross-correlation}

The angular cross-correlation function 
$w_{\rm cross}(\theta)$ is defined as the fractional excess of the 
probability relative to a random distribution \citep{Peebles80},
and can be measured from galaxy samples by the $pair\ counts$ method. 
There are several kinds of estimators that are proposed to measure 
the cross-correlation (e.g. Blake et al.~2006); the one we adopt here is the modified
Landy-Szalay estimator which is derived from the auto-correlation
\citep{Landy93},
\be
w_{\rm cross}(\theta) = \frac{D_1D_2-D_1R_2-D_2R_1+R_1R_2}{R_1R_2},
\ee
where $D_1D_2(\theta)$, $D_1R_2(\theta)$, $D_2R_1(\theta)$ and 
$R_1R_2(\theta)$ are the normalized pair counts for data ($D_i$) and random ($R_i$) catalogs
with separation $\theta$. 

\begin{figure*}
\centerline{
\resizebox{!}{!}{\includegraphics[scale=0.3]{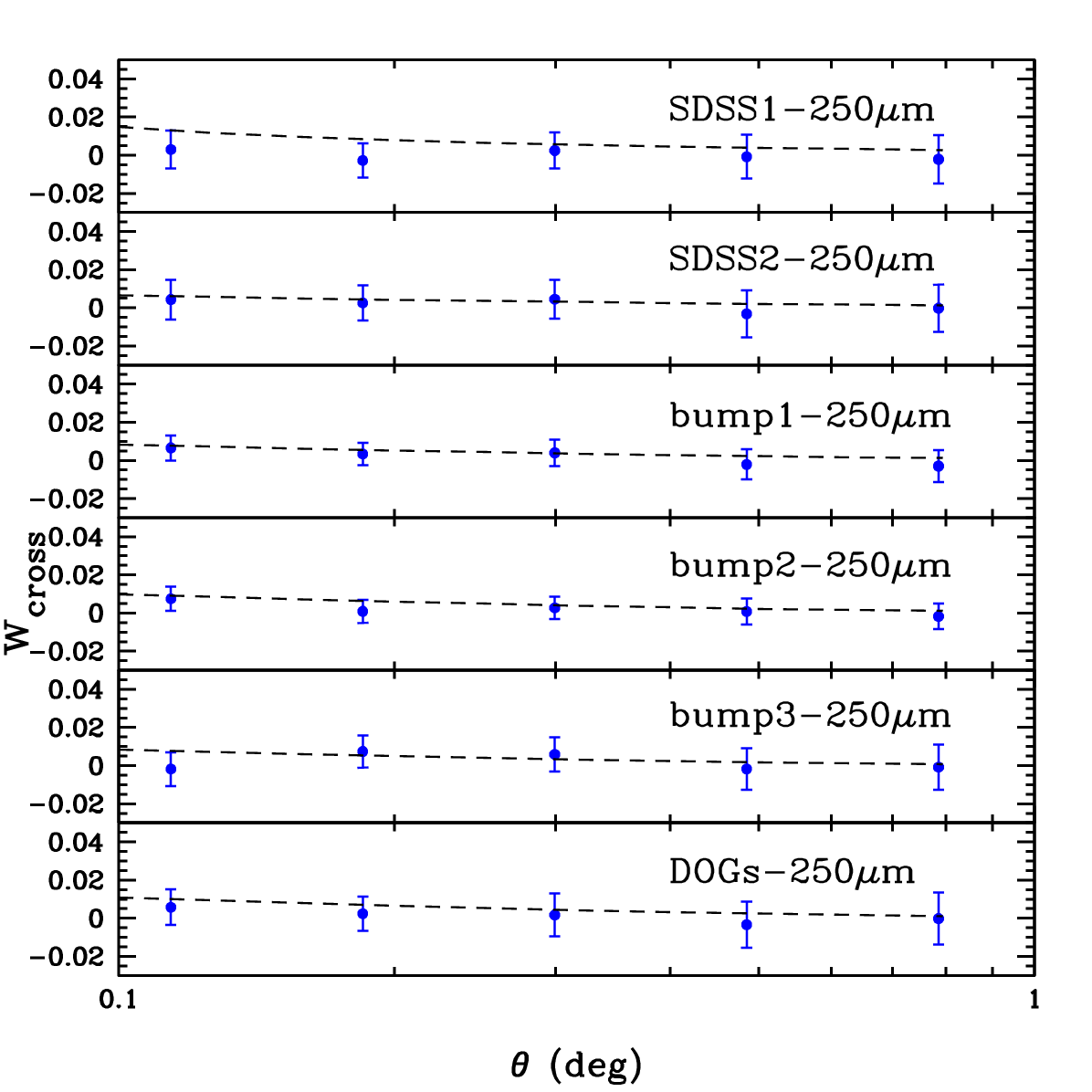}}
\resizebox{!}{!}{\includegraphics[scale=0.3]{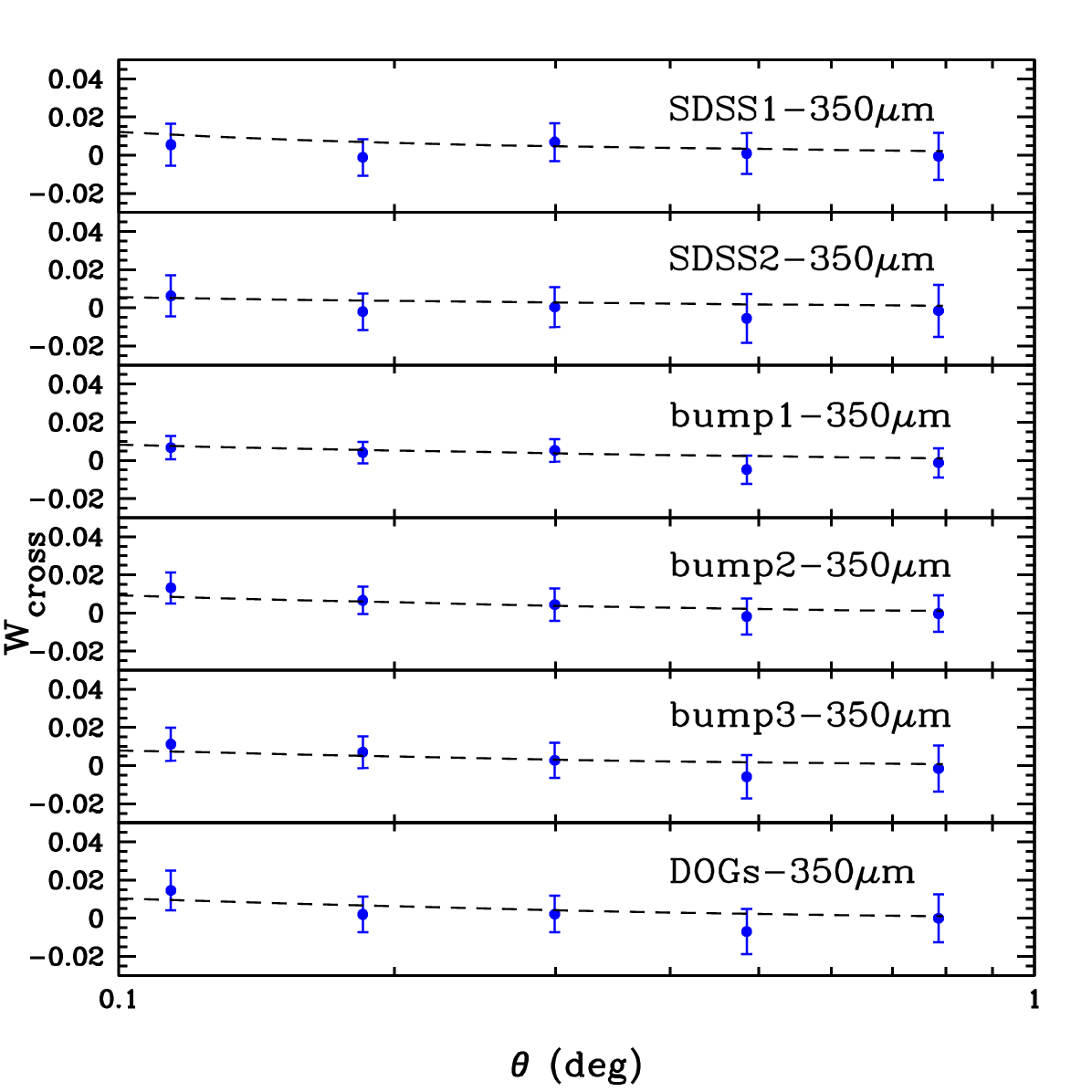}}
\resizebox{!}{!}{\includegraphics[scale=0.3]{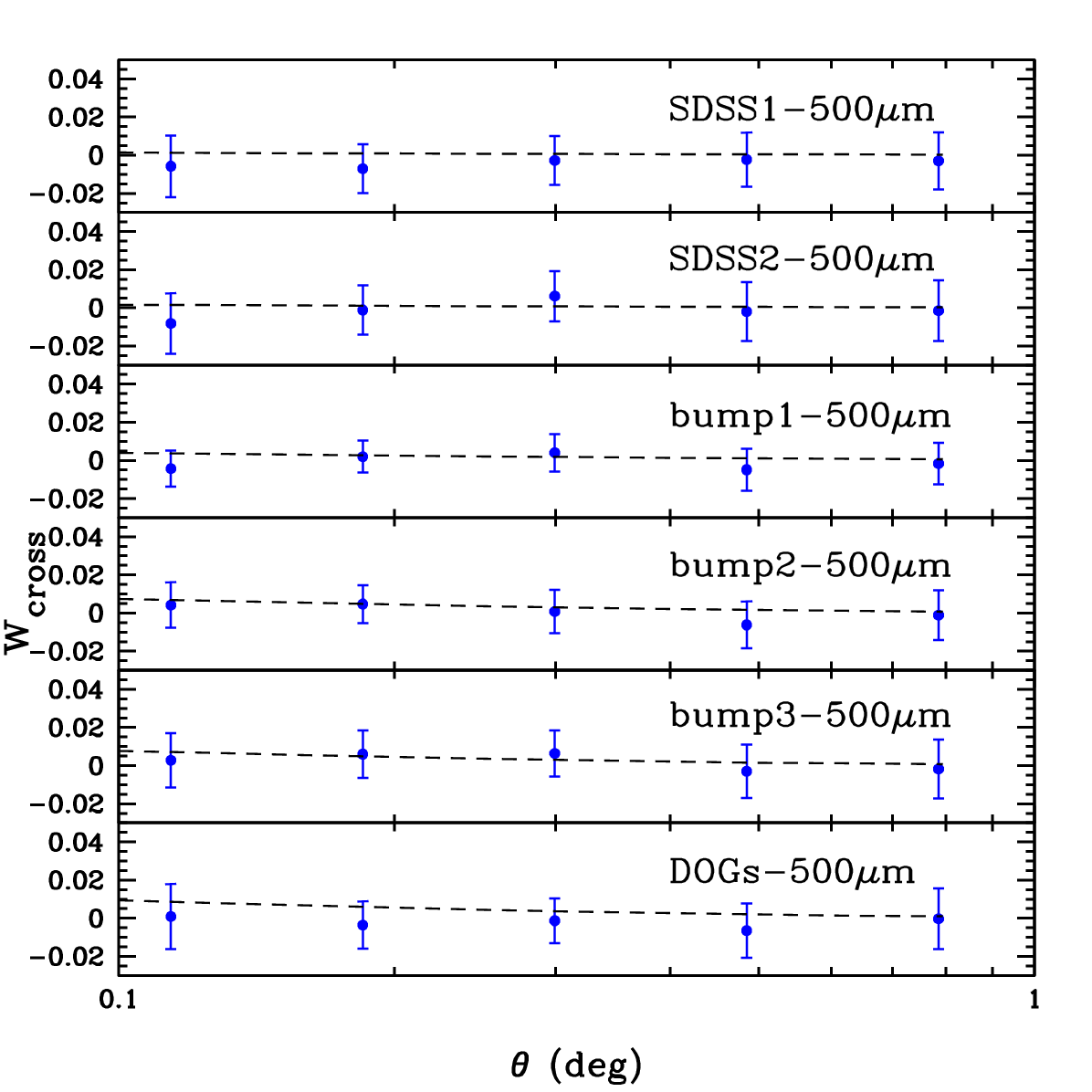}}
}
\caption{Angular cross-correlation between the 250,
350, and 500 $\mu$m objects of the SPIRE surveys and the 
SDSS1, SDSS2, bump1, bump2, bump3 and DOGs in the Bo\"{o}tes field. The
$1\sigma$ error bars are derived from 200 bootstrap realizations.
The black dashed lines are theoretical estimates of the angular 
cross-correlation using the best-fit value of the SPIRE redshift
distribution.
}
\label{fig:w_bsd}
\end{figure*}

\begin{figure*}[htpb]
\centerline{
\resizebox{!}{!}{\includegraphics[scale=0.3]{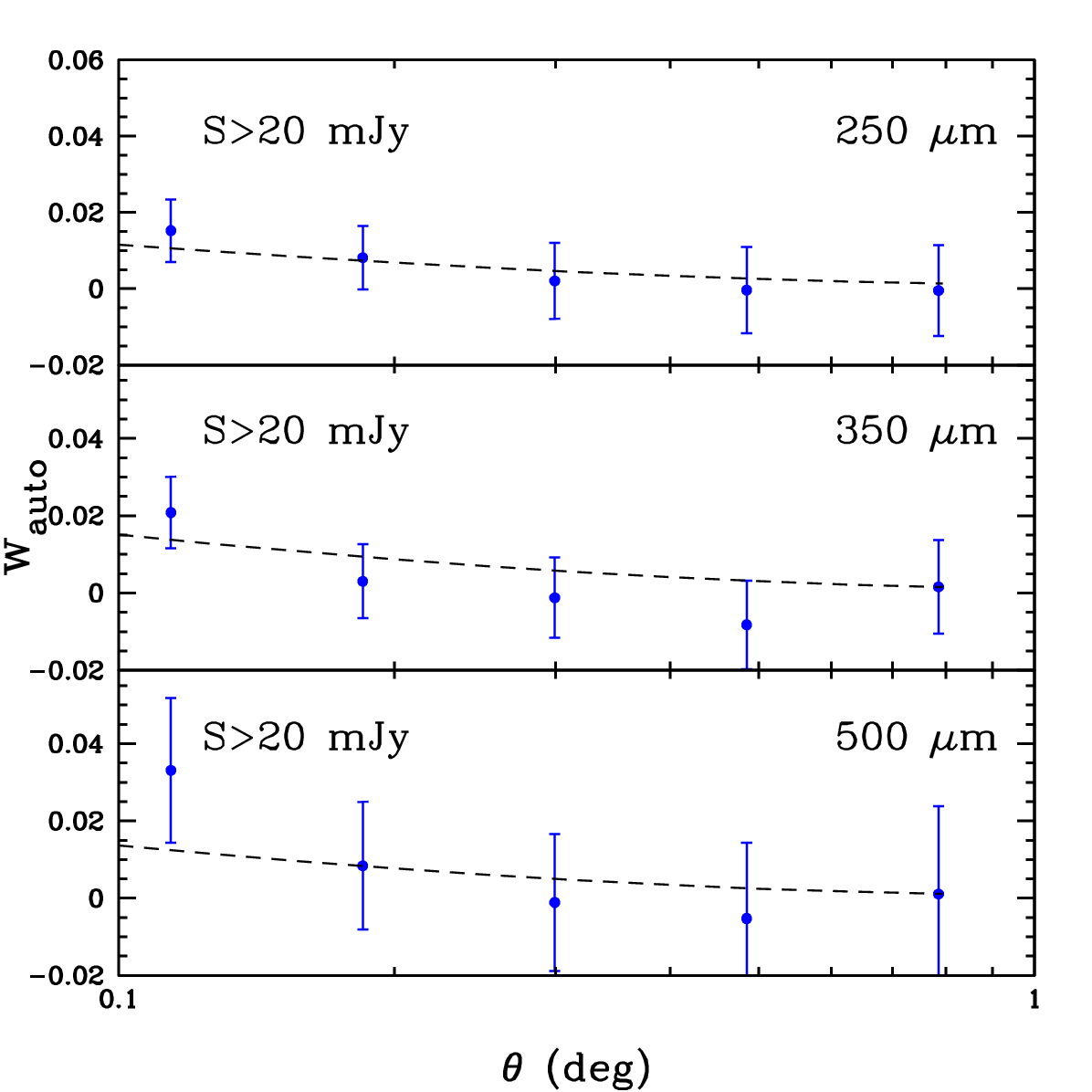}}
\resizebox{!}{!}{\includegraphics[scale=0.3]{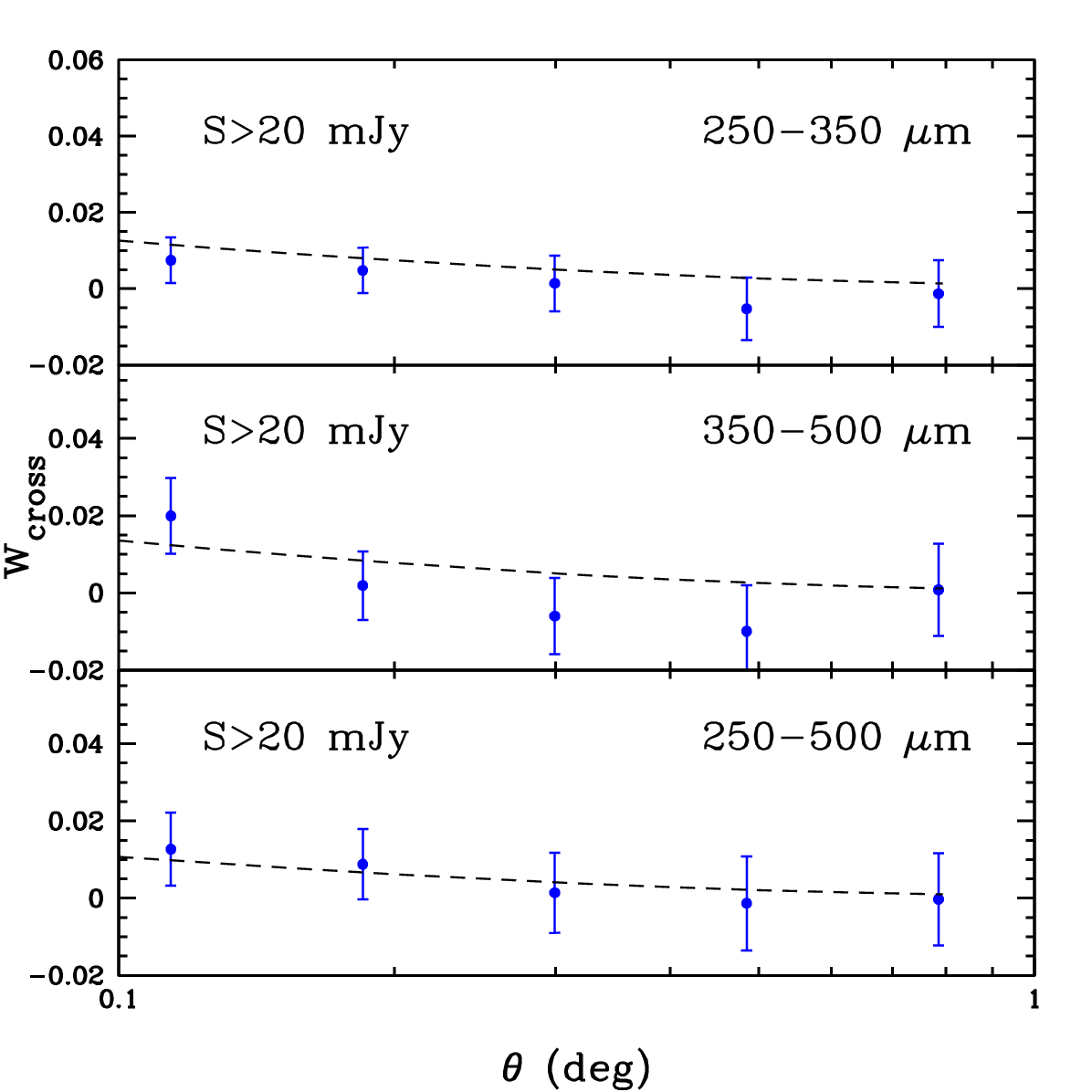}}
}
\caption{Angular auto- and cross-correlation for the 250,
350, and 500$\,\mu$m SPIRE sources in the Bo\"{o}tes field. 
The $1\sigma$ error bars are derived from 200 bootstrap realizations.
The black dashed lines are the theoretical estimation using the best-fit 
value of the SPIRE redshift distribution.
}\label{fig:w_s}
\end{figure*}

We generate random un-clustered catalogs with varying catalog sizes that contain
5 to 10 times more sources than the observed  samples, with a larger number of sources than in data catalogs to avoid biases coming from Poisson fluctuations.  The angular cross-correlation extracted 
from the observational data, and the theoretical estimation using the
best-fit value (see next section) of the SPIRE distribution 
in the Bo\"{o}tes field are shown in Fig.\ref{fig:w_bsd}.
The auto- and cross-correlation of the SPIRE surveys for  250$\,\mu$m,
350$\,\mu$m and 500$\,\mu$m are also
shown in Fig.~\ref{fig:w_s}. 

As was mentioned in the previous Section, to avoid 
biases coming from non-linear clustering
we only use $w(\theta)$ data from 0.1 to 1$^{\circ}$ to fit
the model, since adding the 1-halo term with three or four extra 
parameters for the halo occupation number will result in
extra degeneracies, degrading the $N(z)$ estimates, 
consistent with theoretical suggestions in the literature (e.g., Neyrinck et al.~2006).
Also keeping to scales larger than 0.1$^{\circ}$, we  avoid the need to introduce a 
transfer function for $w(\theta)$ for SPIRE sources
and their cross-correlations since at smallest scales close to the SPIRE point response 
function, clustering is expected to be affected by source blending
and issues related to map making. As studied in Cooray et al. (2010), at $\theta > 
0.05^{\circ}$, there are no corrections to the measured $w(\theta)$.

To evaluate the covariance matrix of the angular correlation $w(\theta)$,
we use a bootstrap method to generate 200 realizations for the
galaxy samples. Then the covariance 
matrix of $w_{\rm cross}$ is
\be
C_{ij} = \frac{1}{N-1}\sum_k^N[w_{k}(\theta_i)-\bar{w}(\theta_i)][w_{k}(\theta_j)-\bar{w}(\theta_j)],
\ee
where $N=200$ is the number of the bootstrap realization,
and $\bar{w}(\theta)$ is the average angular correlation
for all bootstrap realizations at $\theta$. The error of the
angular correlation thus takes the form of
$\sigma_w(\theta_i)=\sqrt{C_{ii}}$.

We use nine logarithmic bins from 0.01 to 1$^{\circ}$ to calculate
the angular auto- and cross-correlation and their covariance matrix. The model 
correlation and cross-correlation functions, $w^{\rm th}$, are calculated for a given
$N(z)$ and clustering bias factors (described in the next Section), and are compared with 
measurements, $w^{\rm data}$, using the covariance matrix 
from the data. In calculating  $w^{\rm th}$, we make use of the measured $N(z)$ of 
the SDSS, bumps and DOGs, derived in the last section. 

\begin{figure*}[th!]
\centerline{
\resizebox{!}{!}{\includegraphics[scale=0.31]{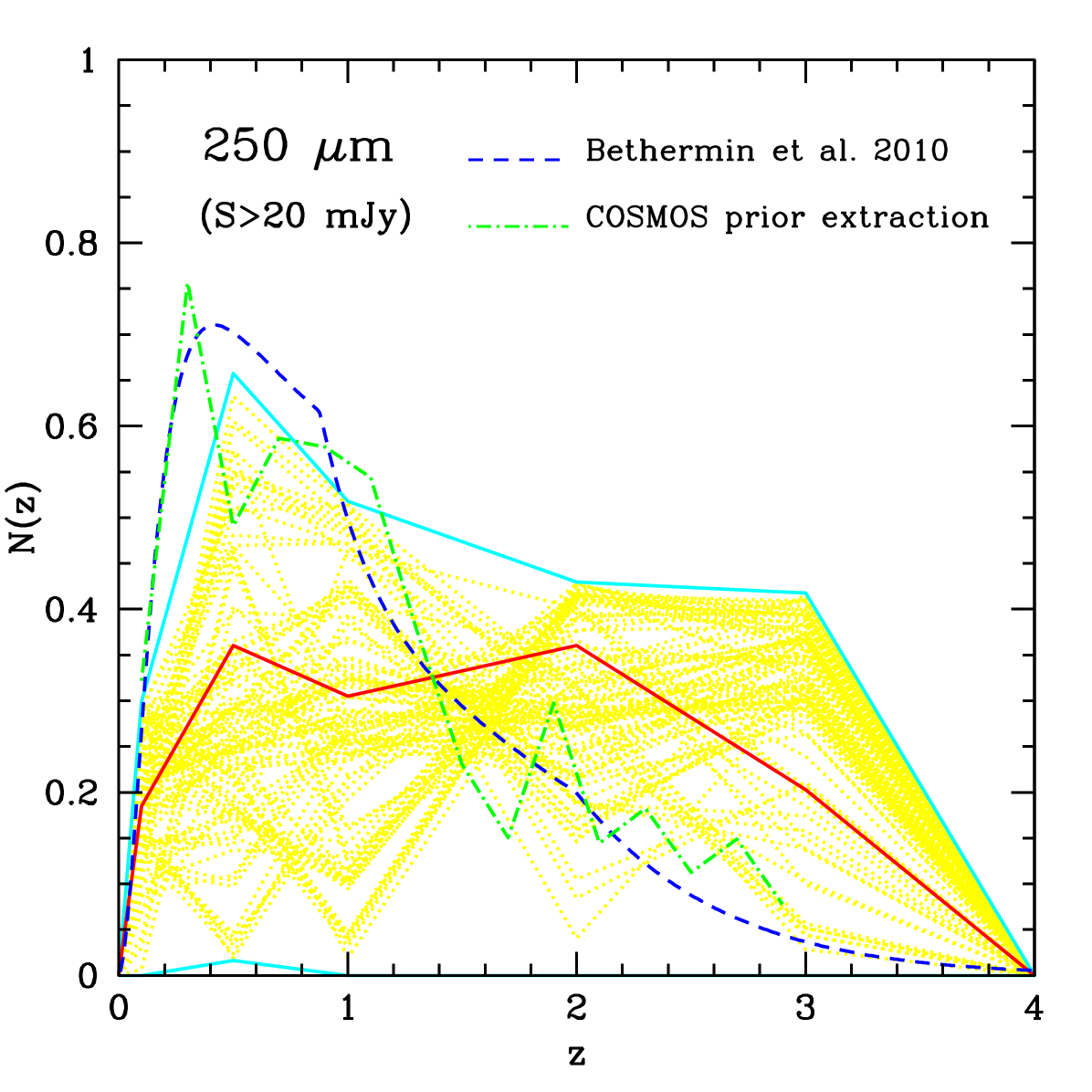}}
\resizebox{!}{!}{\includegraphics[scale=0.31]{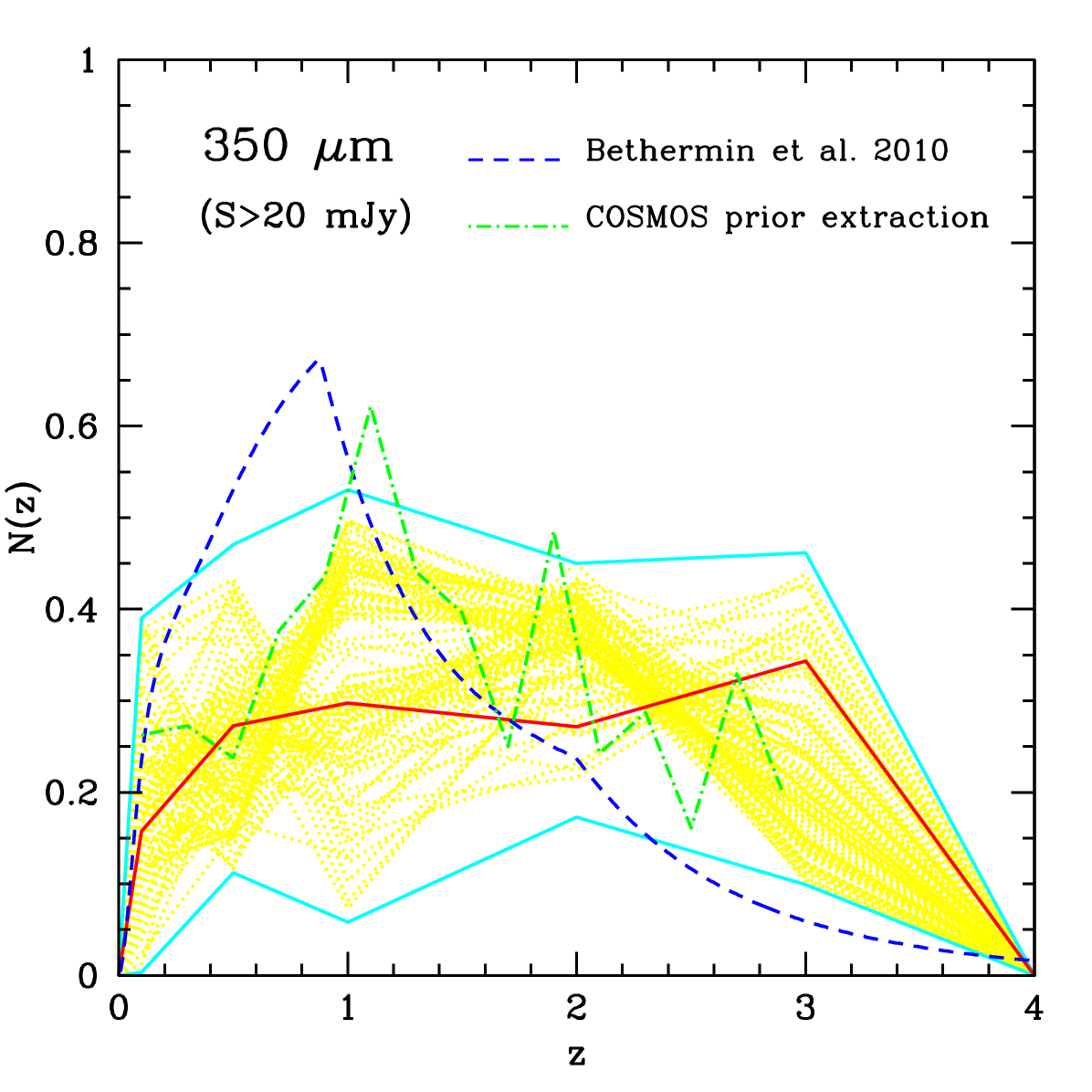}}
\resizebox{!}{!}{\includegraphics[scale=0.31]{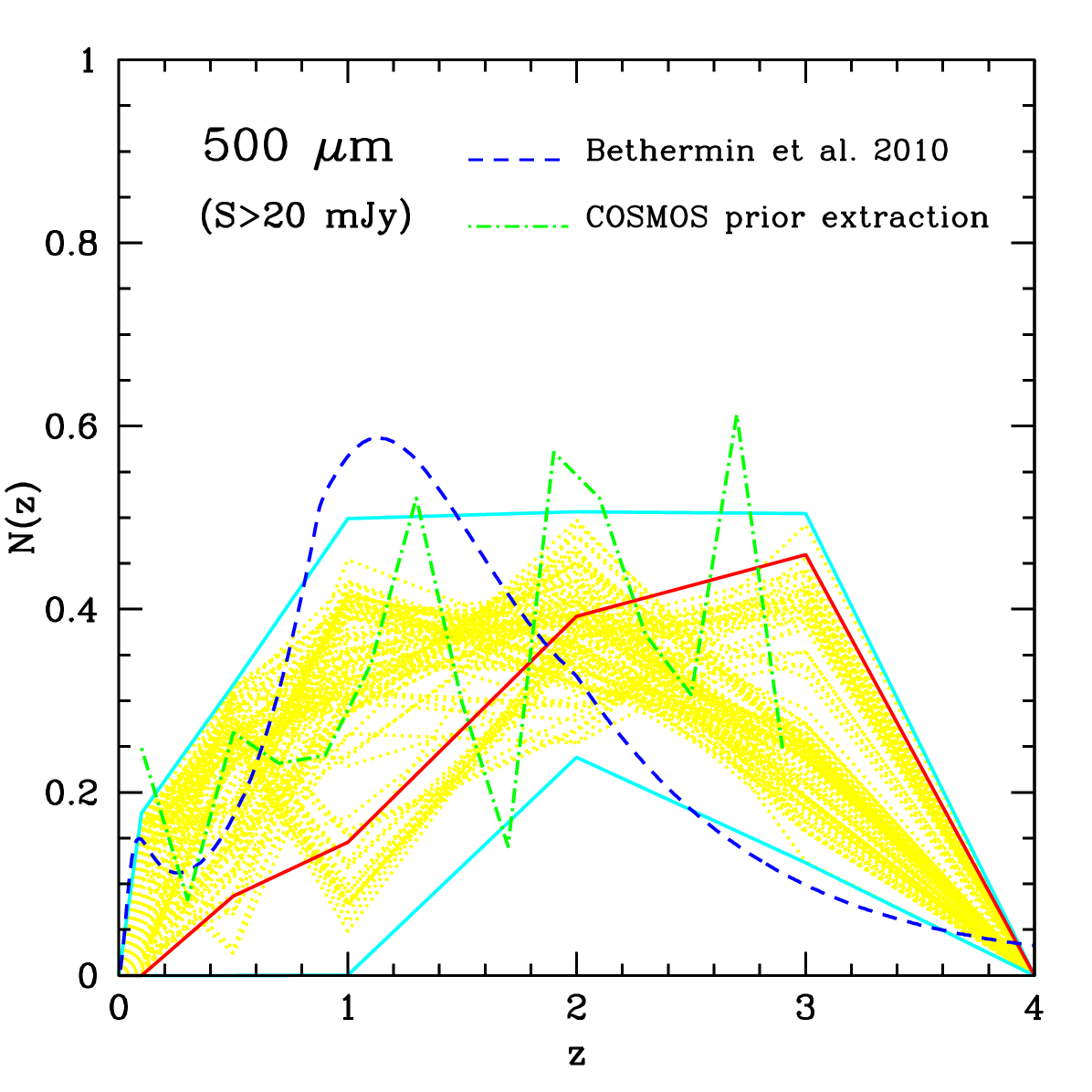}}
}
\caption{Best-fit normalized redshift distributions (red solid line) and 
$1\sigma$ error regions (cyan line) for sources with flux densities greater than 20 mJy for  
250, 350 and 500 $\mu$m SPIRE bands using SDSS, bump-1 to bump-3, and DOGs catalogs in the 
Bo\"{o}tes field. As examples 100 best-fit N(z) from our MCMC results are also shown in yellow dotted lines. 
The analytical model predictions on $N(z)$ from the literature
(blue dashed line; B\'ethermine et al. 2010) for galaxies in the three SPIRE bands are also shown 
for comparison. The green line is a direct estimate of $N(z)$ using the combination of a stacking and 
a cross-identification analysis involving 24 $\mu$m MIPS and SPIRE sources from B\'ethermin et al. (2012).\vspace{.67cm}
}
\label{fig:Nz_SPIRE_25}
\end{figure*}

\section{Estimating the SPIRE galaxy redshift distribution}

We employ a Markov Chain Monte Carlo (MCMC) technique, using the 
Metropolis-Hastings algorithm \citep{Hastings}, to fit
the SPIRE redshift distribution $N(z)$ of sources with 
flux densities greater than 20 mJy at each of the three wavelengths.
We follow established standard procedures in fitting the data, including
thinning of the chains and separation of steps that are part of the initial burn-in period.

We describe the unknown redshift distribution
 $N(z)$ at five values, using five ``pivot'' redshifts $z_{\text{p}}=
0.1, 0.5, 1, 2$ and 3, and set $N(z_{\text{p}}=0)=0$ and $N(z_{\text{p}}=4)=0$
to describe the two end points of the redshift distribution. 
To describe $N(z)$ when $0 < z < 4$, we linearly interpolate the 
fitted $N(z)$ distribution at each of the pivot 
redshifts $z_{\text{p}}$ and use those linearly interpolated values between two pivots
in our model fitting algorithm. The assumption that $N(z>4)=0$ does not 
bias our results since we only expect at most
a few percent of the sub-mm galaxies to be located at $z > 4$ (e.g., Pope \& Chary 2010).
Furthermore, we do not have sensitivity to such high 
redshifts, given that the optical and near-IR galaxy samples we 
have used for the cross-correlation study are restricted to $z < 4$. 

Before deciding on this description, we also considered 
a description of $N(z)$ that involved five bins in redshift, with
$N(z)$ taking the same value in each of the bins. However, we failed to obtain fits to 
the binned case since in the first bin $0 < z < z_1$, $N(z)$ prefers a value
that is non-zero at $z_1$, but zero at $z=0$.  The use of pivot redshifts and linear 
interpolation between pivots avoids the discontinuities that were present with the 
binned case, leading to issues with the numerical integrations of the clustering in  
equation~2.

As discussed earlier (related to equation~2), we also need to 
account for the clustering bias factor of galaxies and SPIRE
sources relative to the linear matter power spectrum. Instead of 
keeping the bias in each of the bins as a free parameter, which leads
to a large number of model parameters to be determined 
from the data, we assume a model for the
galaxy bias, as a function of redshift, to be of the form
\be
b(z) = b_0(1+z)^c,
\ee  
where $b_0$ and $c$ are free parameters to be determined from data using the MCMC analysis.
In addition to this model we also consider two other approaches with: (i) 
$b(z)=b_0+b_1z$, a simple linear interpolation with redshift; and (ii)
$b(z)=b_0$ when $z < 2$ and $b(z) = b_1$ when $z > 2$. We found results 
consistent within 1$\sigma$ uncertainties in both $N(z)$ and $b(z)$ with the power-law form 
when using the linear relation.

For optical and IR galaxy samples we assume that each has an average bias factor, 
and we do not account for the redshift evolution of the bias
factor in each of the galaxy samples. This is a fair assumption 
since each of the samples we have created has a narrow 
redshift distribution compared to the distribution expected for the SPIRE galaxies.

Altogether we have thirteen free parameters in our MCMC fitting,
which contains five parameters for the SPIRE redshift distribution and
six bias parameters for SDSS-1, SDSS-2, bump-1, bump-2, bump-3, and DOGs,
plus two parameters to describe the SPIRE galaxy bias and its evolution with redshift.
While the redshift  distribution and bias factor and evolution for 
SPIRE sources are different at each of the three
SPIRE wavelengths, the bias factors for optical and IR-selected 
galaxies remain the same. Thus, the six bias parameters for the
galaxy samples, with assumed or known redshifts, can be 
determined jointly from cross-correlation data at the three SPIRE wavelengths together with their
auto-correlation functions. We fix all the other cosmological
parameters and assume the flat $\rm \Lambda CDM$ model as mentioned
in Section~1.

We fit the data following the $\chi^2$ distribution estimated as
\be
\chi^2 = \sum_{\rm datasets} {\rm \bf \Delta^T C^{-1} \Delta},
\ee 
where ${\rm \bf \Delta}=[w^{\rm data}(\theta_1)-
w^{\rm th}(\theta_1),\ ...,\ w^{\rm data}
(\theta_9)-w^{\rm th}(\theta_9)]$, $\rm {\bf C}$ is
the covariance matrix of $w(\theta), \,w^{\rm th}$ is obtained
directly from the $N(z)$, and ``data'' 
here are the full angular cross-correlations for the SPIRE, 
SDSS-1, SDSS-2, bump-1, bump-2, bump-3 and DOG samples (21 cross-correlations 
for each SPIRE band), and their auto-correlations in 
the Bo\"{o}tes field. The angular auto-correlation and the cross-correlation between the
SPIRE and the SDSS1, SDSS2, bump1, bump2, bump3 and DOG sub-samples
extracted from the observational data in the Bo\"{o}tes field are shown 
in Figs. \ref{fig:w_bsd} and \ref{fig:w_s} as examples.

We adopt an adaptive step-size Gaussian sampler
given by \cite{Doran04} for the MCMC fitting process. 
The convergence criterion we
take is discussed in \cite{Gelman92}. We generate six chains
with about $10^5$  points after the convergence process. At the end we resample the chains
to get about 10\,000 points to illustrate the probability distribution of the parameters.

\section{Results and Discussion}

In Fig.~\ref{fig:Nz_SPIRE_25} we show the best-fit results and the 
$1\sigma$ errors of the redshift distribution, $N(z)$, for the 
three SPIRE bands (see also Table~1 for the values). The redshift 
distributions are normalized such that $\int dz N(z)=1$.
The $1\sigma$ error bars (cyan lines) are derived from the 
Markov chains, which are statistically estimated via the values 
of $N(z)$ calculated using each chain point at different redshifts. 
As an example, the 100 best-fit N(z) are also
shown in yellow dotted lines. As shown by the errors of $N(z)$ 
at high redshift ($z>3$), the galaxy distribution could be 
larger when going from the $250\,\mu m$ to $500\,\mu m$ bands, 
which implies there may be more high-redshift galaxies for 
the $500\,\mu m$ band than the $250\,\mu m$ and $350\,\mu m$ bands.
In Table~1 we also tabulate the average redshift of the SPIRE sources by calculating $\int dz\, zN(z)$, and 
these values range from $1.8 \pm 0.2$ at 250$\,\mu$m to 1.9 $\pm$ 0.2 for 500$\,\mu$m.
We also derive the correlation coefficient for the $N(z)$ at five pivot 
redshifts from our Marcov chains (see Appendix).
We find the correlation is weak ($\sim$10\%) between adjacent $N_{\rm i}$ pivots for each SPIRE band.

Two additional $N(z)$ predictions from the literature are also shown in the plot
for comparison. The dashed line is a direct estimate of $N(z)$ from PSF-fitted extraction 
using 24 $\mu$m positions as a prior \citep{Beth12} and 
the green curve is a model prediction for the SPIRE redshift distribution \citep{Bethermin10}. 
Our estimation for $N(z)$ for the 250$\,\mu$m band agree 
well with both the direct extraction based on 24$\,\mu$m identifications
and a model prediction, while we find some differences at 350 and 
500$\,\mu$m. However, given the large uncertainties in our
binned $N(z)$ estimate these differences are statistically insignificant.

\begin{figure*}
\centerline{
\includegraphics[scale = .95]{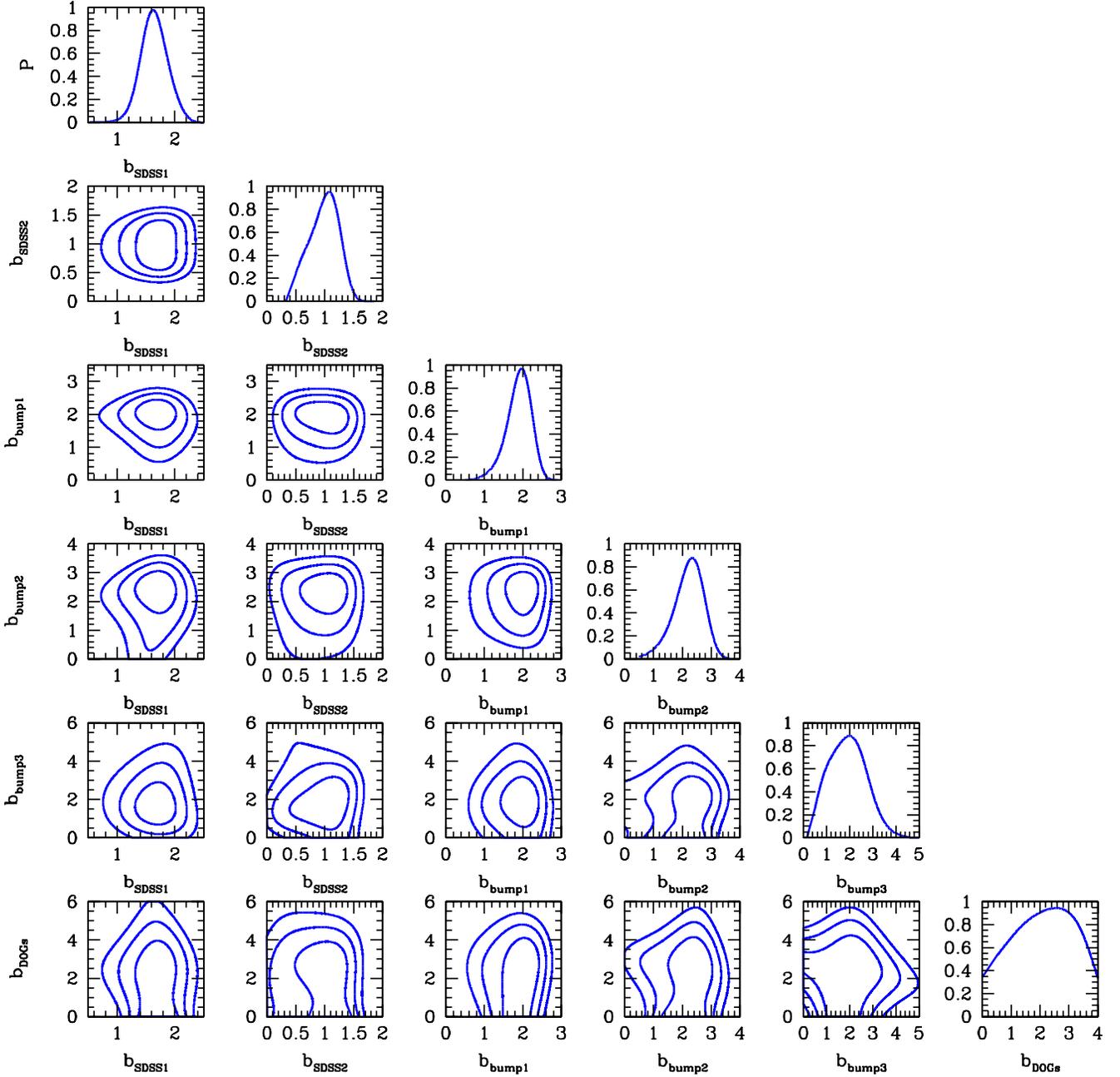}
}
\caption{The one and two-dimensional probability distribution functions for bias parameters
for all of the Bo\"{o}tes samples used throughout this paper. These bias parameters are estimated by
combining the likelihoods from the MCMC chains of all three SPIRE bands.  The 68.3$\%$, 95.5$\%$ and 99.7$\%$ uncertainties from the fits are shown in the two-dimensional error plots.}
\label{fig:bias_fit}
\end{figure*}

\begin{figure}
\centerline{
\includegraphics[scale = 0.4]{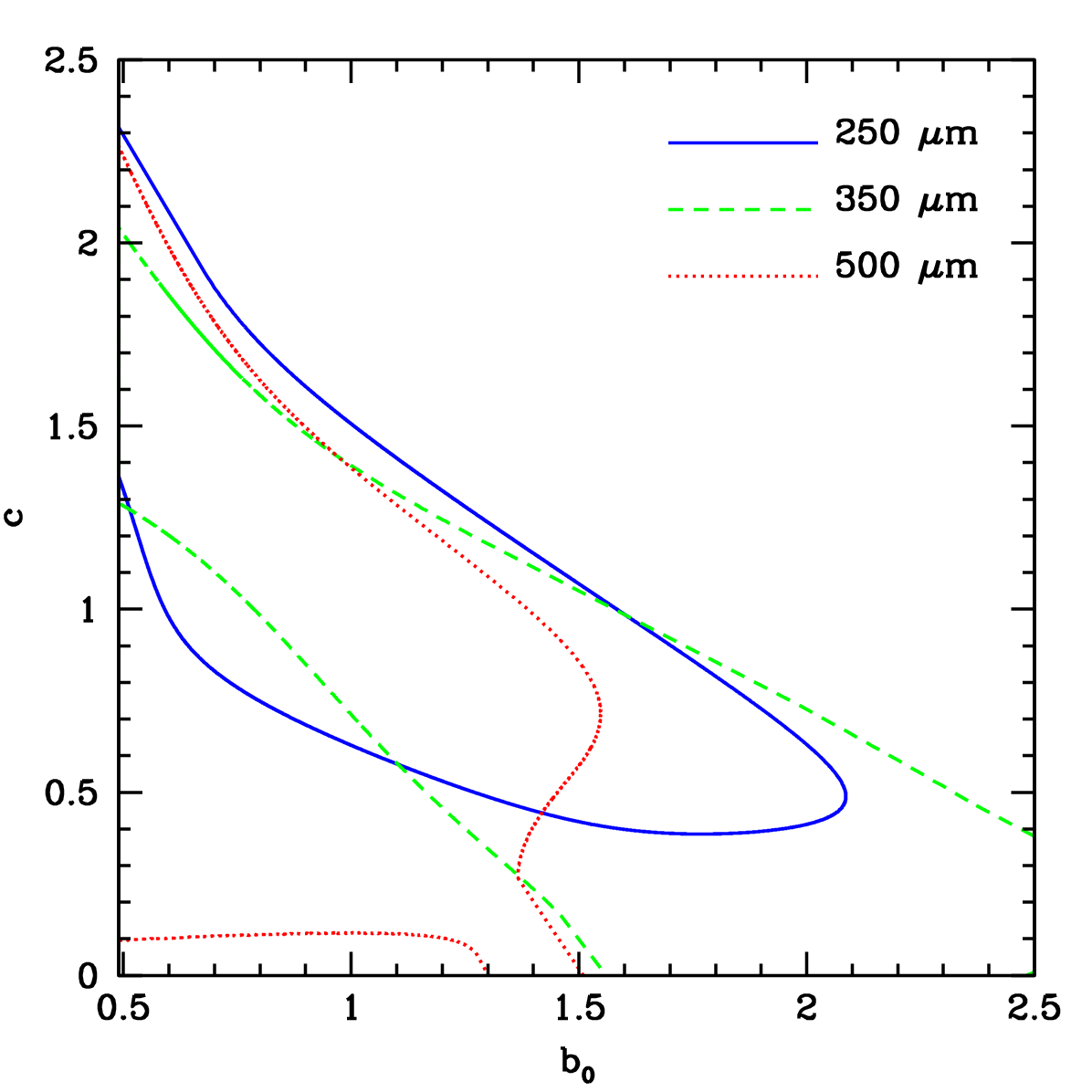}
}
\caption{The 68$\%$ contour maps of the
bias parameters $b_0$ and $c$, determined from the 
MCMC analysis, at 250, 350  and 500$\,\mu$m with
$S>20\,mJy$ in the Bo\"{o}tes field.}
\label{fig:b0_c}
\end{figure}

\begin{figure}
\centerline{
\includegraphics[scale =0.4]{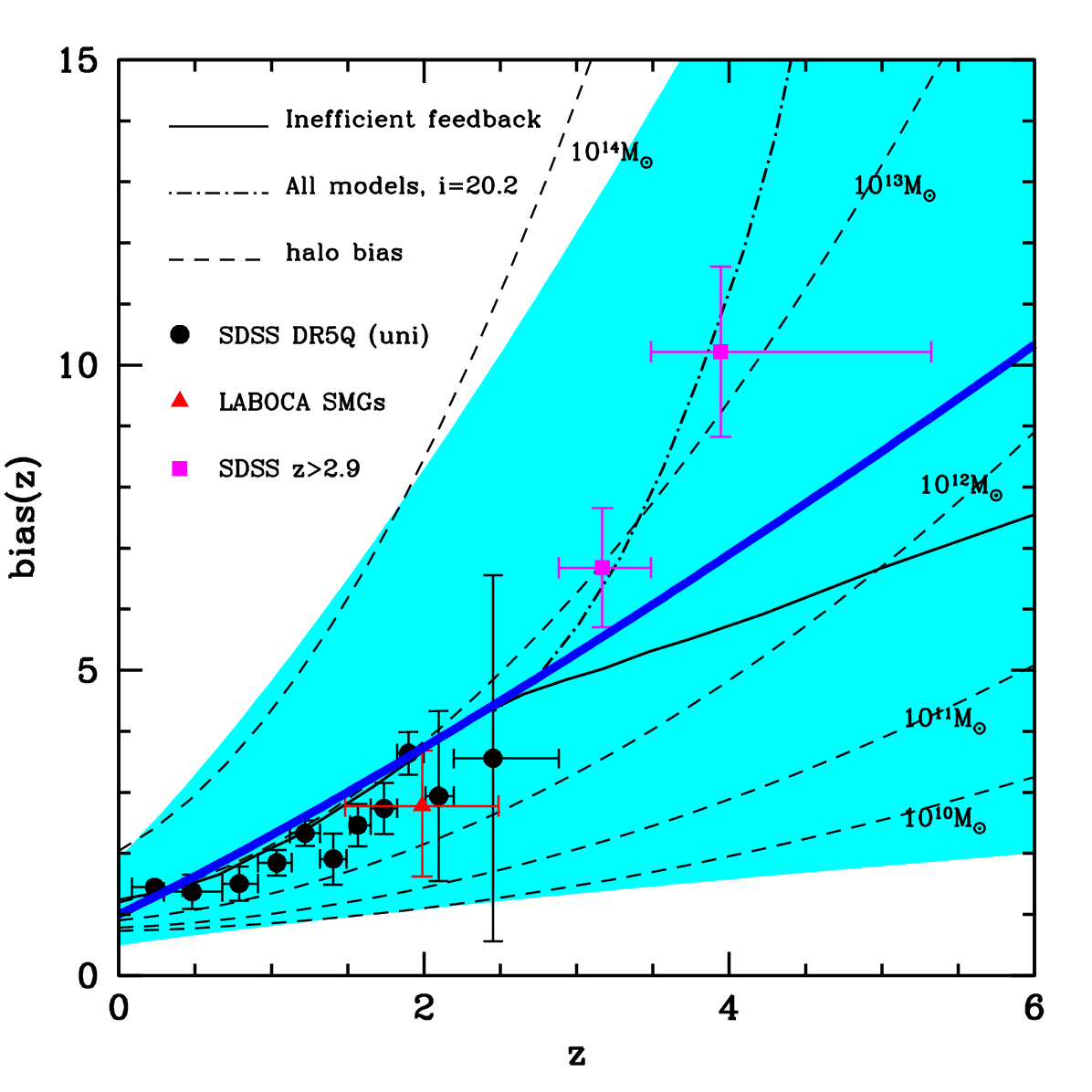}
}
\caption{Clustering bias of $S_{350}>20$ mJy SPIRE sources as a function of redshift. The shaded region 
shows the 68\% confidence region allowed,
with the blue solid line showing the best-fit $b(z)$ relation. For reference we plot the bias factor of 
dark matter halos as a function of halo mass. The range allowed by the
data over $0 < z < 4$ is occupied by halos with mass $10 < \log M/{\text{M}}_{\sun} < 14$. 
We also show samples of galaxies and quasars from the literature \citep{Shen07,Ross09,Hickox12},
and two models for the evolution of the bias factor of merging galaxies from \cite{hopkins} involving
all three models at $i=20.2$ (dash-dotted lines) and inefficient (black solid) feedback.
}
\label{fig:biasz}
\end{figure}

\begin{table}
\centering                    
\caption{The best-fit SPIRE redshift distribution and bias parameters}
\begin{tabular}{c c c c c}
\hline\hline
$N(z)$  & $z$-pivot & $250\ \rm \mu m$ & $350\ \rm \mu m$ & $500\ \rm \mu m$ \\
\hline \\
$N_1$ & 0.1 & $0.19^{+0.11}_{-0.19}$ & $0.16^{+0.23}_{-0.15}$ & $0.00^{+0.18}_{-0.00}$ \\ \\
$N_2$ & 0.5 & $0.36^{+0.30}_{-0.34}$ & $0.27^{+0.20}_{-0.16}$ & $0.09^{+0.23}_{-0.09}$ \\ \\
$N_3$ & 1.0 & $0.31^{+0.21}_{-0.31}$ & $0.30^{+0.23}_{-0.24}$ & $0.15^{+0.35}_{-0.15}$ \\ \\
$N_4$ & 2.0 & $0.36^{+0.07}_{-0.36}$ & $0.27^{+0.18}_{-0.10}$ & $0.39^{+0.11}_{-0.15}$ \\ \\
$N_5$ & 3.0 & $0.20^{+0.22}_{-0.20}$ & $0.34^{+0.12}_{-0.24}$ & $0.46^{+0.04}_{-0.34}$ \vspace{.07cm}\\
\hline
Average Redshift & & & & \\
$\langle z \rangle$   &  & 1.8 $\pm 0.2$ &  1.9 $\pm 0.2$ & 1.9 $\pm 0.2$ \\ \\
\hline
Sub-mm Bias & & & & \\
\hline \\
$b_0$ & & $1.0^{+0.8}_{-0.5}$ & $1.0^{+1.0}_{-0.5}$ & $0.9^{+0.6}_{-0.5}$ \\ \\
$c$ & & $1.1^{+0.4}_{-0.6}$ & $1.2^{+0.3}_{-0.7}$ & $1.1^{+0.5}_{-0.8}$ \vspace{.07cm} \\ 
\hline \hline
\end{tabular}
\label{tab:Nz_bias}
\end{table}

\begin{table}[htbp]
\centering                    
\caption{Bias factors of optical and IR-selected galaxy samples}
\begin{tabular}{c c c}        
\hline\hline       
Sample & Approximate $z$-Range & bias \\
\hline \\
SDSS-1 & $0-0.4$ & $1.6^{+0.2}_{-0.2}$ \\ \\
SDSS-2 & $0.3-0.7$ & $1.1^{+0.2}_{-0.3}$ \\ \\
Bump-1 & $0.8-1.5$ & $2.0^{+0.3}_{-0.3}$ \\ \\
Bump-2 & $1.2-2.0$ & $2.3^{+0.4}_{-0.5}$ \\ \\
Bump-3 & $1.6-2.5$ & $2.0^{+0.8}_{-1.1}$ \\ \\
DOGs & $0.7-3.0$ & $2.6^{+1.1}_{-1.9}$ \vspace{.15cm}\\ 
\hline \hline                   
\end{tabular}
\label{tab:bias}     
\end{table}

In addition to $N(z)$ and bias factors of SPIRE-selected galaxies, 
we also measure the bias factors of the optical and IR-selected galaxy samples that
we have used for cross-correlations.  In Fig.~\ref{fig:bias_fit} we show the two-dimensional error plots, and in
Table~\ref{tab:bias} we list the best-fit bias values and their uncertainties.
These results are obtained by combining the likelihoods from the MCMC chains of all three SPIRE bands. 
These values are consistent with values quoted in the literature for the bias of these samples. For example,
the dust-obscured galaxies have an estimated bias factor of 
$3.1 \pm 0.5$ (Brodwin et al. 2008), which can be compared to our
estimate of $2.6^{+1.1}_{-1.9}$. While fully consistent with the Brodwin 
et al. (2008) estimate, our central value is lower than
their value, as we account for the full redshift distribution of these galaxies, 
while their analysis assumed a redshift of 2 for the whole
DOG sample in the Bo\"otes field.

In Fig.~\ref{fig:b0_c}  we show the 68$\%$ confidence contour maps
of the bias factors of SPIRE sources at the three wavelengths (the values and uncertainties are listed in Table~1). We generally find that the
SPIRE galaxy bias factors are consistent with $b(z)\sim 1+z$ (i.e. 
c$\,\approx\,$1). To understand further the evolution of the sub-mm galaxy bias factor,
we plot the redshift dependence in Fig.~\ref{fig:biasz}, where we compare with 
the bias factor of dark matter halos at several halo masses, from dwarf
galaxy mass to galaxy cluster scales.
The bias factors we find at all three wavelengths indicate a halo mass 
in the range of few times 10$^{10}$ to few times 10$^{13}$ M$_{\sun}$.
The SPIRE clustering analysis in Cooray et al. (2010) found a halo mass for 
sub-mm galaxies that is about 3$\times10^{12}$~M$_{\sun}$, under the assumption of a redshift distribution
for the sub-mm galaxy population with a peak at $z \sim 2.3$, similar to the 
DOG redshift distribution in Fig.~2. We now find a slightly
lower bias factor, and this is primarily due to the fact that the underlying 
redshift distribution of the SPIRE galaxies, especially at 250 $\mu$m, contains more
sources at lower redshifts $(z \lesssim 1)$. While the result here is for bright 
sub-mm sources that are individually detected, the model interpretation of the
SPIRE anisotropy power spectrum by Amblard et al.~(2011) found a minimum 
halo mass of $3 \times 10^{11}\,\text{M}_{\sun}$.

In Fig.~\ref{fig:biasz} we also compare the SPIRE sub-mm galaxy bias 
factors to samples of galaxies and quasars from the literature \citep{Shen07,Ross09,Hickox12}. 
Our results are generally consistent with the possibility that SMGs and 
quasars trace similar evolutionary paths and that the hosts correspond to
dark matter halos that contain a $\sim$ few {\it L}$_{\star}$ ellipticals at 
$z \sim 0$. The exact mechanism on how the starburst galaxies
seen in SPIRE feed the black holes that result in the quasars, and the 
subsequent feedback that suppresses star-formation, remains uncertain.

In Fig.~\ref{fig:biasz} we also plot two models for the evolution of the 
bias factor of merging galaxies from Hopkins et al. (2007).
While these models have similar behavior at $z < 3$, differences 
exist at higher redshift. A clustering study of SPIRE-selected sub-mm galaxies
at $z > 4$ on its own, or as a cross-correlation with high-redshift 
quasars, could potentially be used to understand the intricate role
of starbursts and quasars and to separate the subsequent feedback processes.

\section{Conclusions}

The wide-area sub-mm surveys with the SPIRE instrument aboard the {\it Herschel} Space Observatory 
have now led to catalogs of order one hundred thousand dusty, star-forming galaxies at 250, 350, and 500 $\mu$m.
While some properties of this sub-mm source population are now understood, 
the redshift distribution of these galaxies, $N(z)$,
is not yet well determined observationally. We make a statistical estimate of 
$N(z)$ using a clustering analysis involving the cross-correlation
of sub-mm galaxies detected at each of 250, 350 and 500$\,\mu$m  from 
the {\it Herschel} Multi-tiered Extragalactic Survey (HerMES)
centered on the Bo\"{o}tes field, against samples of galaxies detected at 
optical and near-IR wavelengths from the Sloan Digital Sky Survey (SDSS),
the NOAO Deep Wide Field Survey (NDWFS), and the {\it Spitzer} Deep Wide Field Survey (SDWFS).  

We create optical and near-IR galaxy samples
based on their photometric or spectroscopic redshift distributions 
and test the accuracy of these redshift distributions with similar galaxy
samples defined via catalogs of the Cosmological Evolution Survey (COSMOS). 
We fit the clustering auto and cross-correlations of SPIRE and optical/IR galaxy samples
at angular scales of 0.1 to 1$^{\circ}$, where clustering of each of the galaxy samples 
is expected to be linear, with the amplitude determined by a bias 
factor together with the redshift distribution of the sources.
We make use of a Markov Chain Monte Carlo (MCMC) method to sample $N(z)$ at 
five nodes in the range $0 < z <4$, as well as the bias factors.
The SPIRE-selected sub-mm galaxy bias factor is found to 
vary with redshift according to $b(z)=1.0^{+1.0}_{-0.5}(1+z)^{1.2^{+0.3}_{-0.7}}$.
We find clear evidence of evolving redshift distributions as the 
wavelength increases from 250$\,\mu$m to 500$\,\mu$m, with the 250$\,\mu$m band 
containing the largest number of low redshift sources.
We also compare the measured redshift distribution to model 
predictions in the literature, and find an excess of sources in the 
highest redshift bin when compared to the model prediction from 
B\'ethermin et al. (2010), although in general our results agree with both 
predictions from the literature. With subsequent observations
in more fields, this analysis could potentially be carried out again -- 
incorporating more data in this analysis would reduce the size of the errors and more fully
constrain the $N(z)$ of these sub-mm galaxies.\\

\begin{acknowledgments}
SPIRE has been developed by a consortium of institutes led
by Cardiff Univ. (UK) and including: Univ. Lethbridge (Canada);
NAOC (China); CEA, LAM (France); IFSI, Univ. Padua (Italy);
IAC (Spain); Stockholm Observatory (Sweden); Imperial College
London, RAL, UCL-MSSL, UKATC, Univ. Sussex (UK); and Caltech,
JPL, NHSC, Univ. Colorado (USA). This development has been
supported by national funding agencies: CSA (Canada); NAOC
(China); CEA, CNES, CNRS (France); ASI (Italy); MCINN (Spain);
SNSB (Sweden); STFC, UKSA (UK); and NASA (USA).

The data presented in this paper will be released through the HeDaM Database in Marseille. This work made use of images and/or data products provided by the NOAO Deep Wide-Field Survey, which is supported by the National Optical Astronomy Observatory (NOAO). NOAO is operated by AURA, Inc., under a cooperative agreement with the National Science Foundation. 

Funding for the SDSS and SDSS-II has been provided by the Alfred P. Sloan Foundation, the Participating Institutions, the National Science Foundation, the U.S. Department of Energy, the National Aeronautics and Space Administration, the Japanese Monbukagakusho, the Max Planck Society, and the Higher Education Funding Council for England. The SDSS Web Site is http://www.sdss.org/.
The SDSS is managed by the Astrophysical Research Consortium for the Participating Institutions. The Participating Institutions are the American Museum of Natural History, Astrophysical Institute Potsdam, University of Basel, University of Cambridge, Case Western Reserve University, University of Chicago, Drexel University, Fermilab, the Institute for Advanced Study, the Japan Participation Group, Johns Hopkins University, the Joint Institute for Nuclear Astrophysics, the Kavli Institute for Particle Astrophysics and Cosmology, the Korean Scientist Group, the Chinese Academy of Sciences (LAMOST), Los Alamos National Laboratory, the Max-Planck-Institute for Astronomy (MPIA), the Max-Planck-Institute for Astrophysics (MPA), New Mexico State University, Ohio State University, University of Pittsburgh, University of Portsmouth, Princeton University, the United States Naval Observatory, and the University of Washington.

KMW, YG, AC, JS and JLW are supported by NASA funds for 
US participants in {\it Herschel} through an award from JPL.  
AC and YG also acknowledge support from NSF CAREER AST-0645427 (to AC)
and KMW acknowledges support from a NSF REU supplement.
SO, AS and LW acknowledge support from the Science and Technology Facilities 
Council [grant number ST/F002858/1] and [grant number ST/I000976/1].
AF, LM and MV were supported by the Italian Space Agency (ASI "Herschel Science" 
Contract I/005/07/0).
\end{acknowledgments}

\appendix

\section{The correlation coefficient of the N(z)}

In Table \ref{tab:cov_Nz} we show the correlation coefficient of the N(z) at five pivot redshifts ($N_{\rm i}$) for three SPIRE bands, 
which is derived from our Marcov chains. The definition is given by 
\be
r=\frac{{\rm cov}(N_{\rm i},N_{\rm j})}{\sigma_{N_{\rm i}}\sigma_{N_{\rm j}}}.
\ee
Here ${\rm cov}(N_{\rm i},N_{\rm j})$, $\sigma_{N_{\rm i}}$ and $\sigma_{N_{\rm j}}$ are the covariance matrix and 
standard deviations for $N_{\rm i}$ and $N_{\rm j}$, respectively.

\begin{table}[h!tbp]
\centering                    
\caption{The correlation coefficient of the N(z$_i$) at five pivot redshifts for three SPIRE bands.}
\begin{tabular}{c| c c c c c| c c c c c| c c c c c}        
\hline\hline       
& &  &250 $\rm \mu m$&  &  &  &  & 350 $\rm \mu m$ &  &  &  &  & 500 $\rm \mu m$ &  & \\
& N$_1$ & N$_2$ & N$_3$ & N$_4$ & N$_5$ & N$_1$ & N$_2$ & N$_3$ & N$_4$ & N$_5$ & N$_1$ & N$_2$ & N$_3$ & N$_4$ & N$_5$\\
\hline \hline
N$_1$ & 1.00 &  &  &  &  & 1.00 &  &  &  &  & 1.00 &  &  &  &\\ 
N$_2$ & -0.03 & 1.00 &  &  &  & 0.00 & 1.00 &  &  &  & 0.05 & 1.00 &  &  &\\ 
N$_3$ & -0.02 & -0.18 & 1.00 &  &  & 0.00 & -0.03 & 1.00 &  &  & 0.09 & 0.00 & 1.00 &  &\\ 
N$_4$ & 0.08 & -0.10 & -0.05 & 1.00 &  & -0.11 & -0.09 & -0.05 & 1.00 &  & 0.12 & 0.04 & -0.06 & 1.00 & \\ 
N$_5$ & -0.01 & -0.01 & -0.06 & -0.10 & 1.00 & 0.05 & 0.10 & 0.02 & -0.13 & 1.00 & 0.06 & 0.07 & -0.16 & -0.11 & 1.00\\
\hline \hline                  
\end{tabular}
\label{tab:cov_Nz}     
\end{table}


\begin{thebibliography}{99}
\bibitem[Ashby et al.(2009)]{Ash09} Ashby, M.L.N., et al. 2009, ApJ, 701, 428 
\bibitem[Amblard et al.(2010)]{Amb10} Amblard, A., et al. 2010, A\&A 518, 9L
\bibitem[Abazajian et al. (2009)]{Aba09} Abazajian, et al. 2009, ApJ Supplement Series, 182, 543
\bibitem[Bartelmann \& Schneider(2001)]{Bartelmann01} Bartelmann, M. \& Schneider, P. 2001, Physics Reports, 340, 291
\bibitem[Bessell\& Brett(1988)]{Bes88} Bessell, M.S., \& Brett, J.M. 1988, PASP, 100, 1134
\bibitem[B\'{e}thermin et al.(2010)]{Bethermin10} B\'ethermin, M., Dole, H., Lagache, G., Le Borgne, D. \& Penin, A. 2010, A\&A, 529, A4
\bibitem[B\'{e}thermin et al. (2012)]{Beth12} B\'ethermin, et al. 2012, arXiv:1203.1925
\bibitem[Blake et al.(2006)]{Blake06} Blake, C., Pope, A., Scott, D. \& Mobasher, B. 2006, MNRAS, 368, 732
\bibitem[Browdin et al.(2008)]{Bro08} Browdin, M., et al. 2008, ApJ, 687, L65
\bibitem[Capak et al. (2007)]{Capak07} Capak, P.m et al. 2007, ApJS, 172, 99
\bibitem[Csabai et al. (2003)]{Csa03} Csabai, I., et al. 2003, AJ, 125, 580
\bibitem[Cooray \& Sheth(2002)]{Cooray02} Cooray, A. \& Sheth, R. 2002, Physics Report, 372, 1
\bibitem[Dey(2009)]{Dey09} Dey, A. 2009, ASPC, 408, 411
\bibitem[Dey(2008)]{Dey08} Dey, A., et al. 2008, ApJ, 677, 943
\bibitem[Doran \& Mueller(2004)]{Doran04} Dpran, M. \& Mueller, C. M. 2004, JCAP, 0409, 003
\bibitem[Eisenhardt et al.(2004)]{Eis04} Eisenhardt, P.R., et al. 2004, ApJ, 154, 48
\bibitem[Gong \& Chen(2007)]{Gong07} Gong, Y. \& Chen, X. 2007, \prd, 76, 123007
\bibitem[Gelman \& Rubin(1992)]{Gelman92} Gelman, A. \& Rubin, D. 1992, Stat. Sci., 7, 457
\bibitem[Harris et al.(2012)]{Harris} Harris, A. et al. 2012, ApJ submitted.
\bibitem[Hastings(1970)]{Hastings}Hastings, W.K. "Monte Carlo Sampling Methods Using Markov Chains and Their Applications". {\it Biometrika} 57 (1): 97Ð109. 
\bibitem[Hickox et al.(2012)]{Hickox12} Hickox, R. C., et al. 2012, \mnras, 421, 284
\bibitem[Hopkins et al.(2007)]{hopkins} Hopkins, P. F., Hernquist, L., Cox, T. \& Keres, D. 2008, ApJS, 175, 356
\bibitem[Huang et al.(1997)]{Hua97} Huang, J., Cowie, L.L., Gardner, J.P., Hu, E.M., Songaila, A., \& Wainscoat, R.J. 1997, ApJ, 476, 12
\bibitem[Ilbert et al.(2009)]{Ilb09} Ilbert, O., et al. 2009, ApJ, 690, 1236
\bibitem[Januzzi \& Dey(1999)]{Jan99} Jannuzi, B. T. \& Dey, A. 1999, ASP Conference Series, Vol. 191, p. 111
\bibitem[John(1988)]{Joh88} John T.L. 1988, A\&A, 193, 189
\bibitem[Johnson(1966)]{Joh66} Johnson, H.L. 1966, ARA\&A, 4, 193
\bibitem[Komatsu et al.(2011)]{WMAP7} Komatsu, E., et al. 2011, \apjs, 192, 18
\bibitem[Landy \& Szalay(1993)]{Landy93} Landy, S. D. \& Szalay, A. S. 1993, ApJ, 412, 64
\bibitem[Lupu et al.(2010)]{Lupu10} Lupu, R. E. et al. 2010, arXiv.org:1009.5983
\bibitem[Marsden et al.(2009)]{Mars09} Marsden, G., et al. 2009, ApJ, 707, 1729
\bibitem[Moessner \& Jain(1998)]{Moessner98} Moessner, R. \& Jain, B. 1998, MNRAS, 294, L18
\bibitem[Neyrinck et al.(2006)]{Ney06} Neyrinck, M. C., Szapudi, I. \& Rimes, C. D. 2006, MNRAS, 370, L66
\bibitem[Oliver et al.(2012)]{Oli12} Oliver, S. J. et al. 2012, MNRAS submitted.
\bibitem[Peebles(1980)]{Peebles80} Peebles, P. J. E. 1980, The Large-Scale Structure of the Universe. Princeton Univ. Press, Princeton, NJ
\bibitem[Pilbratt et al. (2010)]{Pil10} Pilbratt, G., et al. 2010, A\&A, 518, L1
\bibitem[Pope \& Chary (2010)]{Pope10} Pope, A. \& Chary, R.-R. 2010, ApJ, 715, L171.
\bibitem[Riechers et al.(2011)]{Rie11} Riechers, D. et al. 2011, ApJ, 733, L12
\bibitem[Ross et al.(2009)]{Ross09} Ross, N. P. et al. 2009, ApJ, 697, 1634
\bibitem[Savage \& Oliver(2007)]{Sav07} Savage, R. S., \& Oliver, S. 2007, ApJ, 661, 1339
\bibitem[Sawicki(2002)]{Saw02} Sawicki, M. 2002, AJ, 124, 3050
\bibitem[Schneider et al.(2010)]{Sch10} Schneider, M., Knox, L., Zhan, H., Connolly, A. 2006, ApJ, 651, 14
\bibitem[Scoville et al.(2006)]{Sco06} Scoville, N. et al. arXiv.org:astro-ph/0612305
\bibitem[Scott et al.(2010)]{Scott} Scott, K. S. et al. 2011, ApJ, 733, 29
\bibitem[Shen et al.(2007)]{Shen07} Shen, Y., et al. 2007, AJ, 133, 2222
\bibitem[Simpson \& Eisenhardt(1999)]{Sim99} Simpson, C., \& Eisenhardt, P. 1999, PASP, 111 691
\bibitem[Smith et al.(2012)]{Smi11} Smith, A.J., et al., 2012, MNRAS, 419, 377
\bibitem[Newman(2008)]{New08} Newman, J. A. 2008, ApJ, 684, 88
\bibitem[Waddington et al.(2007)]{Wad07} Waddington, I., et al. 2007, MNRAS, 381, 1437
\bibitem[Wang et al.(2011)]{Wang11} Wang, L., et al. 2011, 414, 596
\bibitem[Wang et al.(2012)]{Wang12} Wang, L., et al. 2012, in preparation.
\bibitem[Wright \& Fazio(1994)]{Wri94} Wright, E.L., Eisenhardt, P., \& Fazio, G. 1994, BAAS, 26, 893
\bibitem[Zhang et al.(2010)]{Zhang} Zhang, P., Pen, U.-L., Bernstein, G. 2010, MNRAS, 405, 359
\end{thebibliography}
\end{document}